\documentclass{article}

\usepackage{arxiv}
\usepackage[numbers]{natbib}
\usepackage{todonotes, multirow}
\usepackage{amsmath, amssymb}
\usepackage{subcaption}
\usepackage{svg}
\usepackage{wasysym}
\usepackage{hyperref}

\begin{document}
\let\WriteBookmarks\relax
\def\floatpagepagefraction{1}
\def\textpagefraction{.001}

\title{Quantitative Fiber Orientation Analysis of Carbon Fiber Sheet Molding Compounds using Polarization Imaging and X-Ray Computed Tomography}                      

\author{Miro Duhovic \texttt{miro.duhovic@leibniz-ivw.de}\\Alex Keilmann \\Dominic Schommer\\Claudia Redenbach\\Katja Schladitz}
\author{
        %% \And
    Miro Duhovic \\
    Leibniz-Institut für Verbundwerkstoffe \\
    \texttt{miro.duhovic@leibniz-ivw.de} \\
    \And
    Alex Keilmann \\
    University of Kaiserslautern-Landau \\
    \& Fraunhofer ITWM\\
    \And
    Dominic Schommer \\
    Leibniz-Institut für Verbundwerkstoffe \\
    \And
    Claudia Redenbach \\
    University of Kaiserslautern-Landau \\
    \And
    Katja Schladitz \\
    Fraunhofer ITWM \\
    %% \And
    %% Coauthor \\
    %% Affiliation \\
    %% Address \\
    %% \texttt{email} \\
}

\maketitle

\begin{abstract}
For the quality inspection of carbon fiber sheet molding compounds, polarization imaging is a promising alternative to more established methods like computed tomography, since it is cheaper, faster, and provides a larger field-of-view. For uni- and bidirectional carbon fiber reinforced composite materials, 
machine vision cameras with on-chip polarized image sensors have been successfully validated for visualizing fiber orientation. Although this imaging technique is already being applied to multidirectional materials, to our knowledge, it has not yet been validated for such materials. In this paper, fiber orientations obtained by angle of linear polarization images (AOLP) of commercially available pressed carbon fiber sheet molding compound materials are compared with orientations estimated from computed tomography scans. The fiber orientations in the computed tomography images are estimated using the maximal response of anisotropic Gaussian filters and the deviation between orientations estimated from polarization and computed tomography images is calculated. Both imaging methods showed encouraging visual similarity, but also notable numerical differences, which are discussed in depth. Moreover, it is shown that the surface layer fiber orientation is representative of the fiber orientation through the entire specimen.
\end{abstract}
\keywords{Molding compounds \and Anisotropy \and Optical properties/techniques \and Radiography}

\section{Introduction}
Carbon fiber sheet molding compounds (C-SMCs) are a type of composite precursor material that is fiber-reinforced, thermoset-based, and produced in flat sheet form. Due to their versatile production and processability, C-SMCs have become increasingly popular for manufacturing various products, particularly in the automotive and aerospace industries. C-SMCs enable molding of complex shaped geometries resulting in lightweight parts with high stiffness and strength. 
These properties are determined by the C-SMC's microstructure, in particular by the carbon fiber component, and its orientation \cite{fu96, vallons09, yokozeki06}. Therefore, estimating the material's fiber orientation is a vital issue, e.g., for quality inspection \cite{heuer2015review} or numerical simulations of part manufacturing processes \cite{schommer2019material, zhang11}.

\subsection{Methods for Determining Fiber Orientation Distributions in CFRP}

Many non-destructive testing methods can be used to measure fiber orientation in carbon fiber-reinforced plastic (CFRP) parts. In the following, we will specifically discuss X-ray radiography and X-ray computed tomography (CT),  the (electrical) eddy current method, thermography, and optical methods such as confocal laser microscopy and polarization imaging. However, most of them are not suitable for inline quality inspection due to their limited accuracy, speed, or expense.

In X-ray radiography, the sample is penetrated by X-rays yielding a contrast turned into a 2D image. Besides the conventional X-ray absorption contrast, phase contrast or dark-field contrast can also be measured~\cite{denos2018fiber,  garcea2018x, jensen2010directional, jensen2010directionalstrongly, malecki2013x, prade2017nondestructive, sharma2018advanced}. From many X-ray radiographies taken at varying rotation angles, three-dimensional images of the specimen under consideration can be computed by so-called tomographic reconstruction.
For analyzing the microstructure of fiber-reinforced plastics, micro-computed tomography (\textmu CT) with resolutions in the micro-meter range is used. Resolutions down to the fiber level (ideally between 4 and 12\,pixels for a fiber diameter of 6-8\,\textmu m  for carbon fibers and 10\,-\,12\,\textmu m for glass fibers) can be achieved by both, synchrotron radiation based \textmu CT and modern laboratory devices~\cite{schladitz2011quantitative}. The 3D fiber orientation can be deduced from these \textmu CT images to a high level of detail for a wide range of morphologies~\cite{bidola2017application, jiang20203d, latil2011towards, le2008XrayPhaseContrast, martulli2019carbon, sabiston2019method, schladitzNondestructiveCharacterizationFiber2017, sentis20173d, trauth2021effective}, given a sufficiently high gray value contrast between the fiber system and the other components. The latter is however a major challenge for laboratory CT devices as the X-ray absorption contrast of carbon and most polymer matrix materials differs only slightly.
Moreover, in view of in-line application, \textmu CT is still challenged by
rather narrow fields-of-view determined by the detector size and the required spatial resolution. 
Further challenges such as costs, size of the device, and radiation protection rules for operation still limit this technology to research laboratory environments and restrict its use within industrial settings and online measurements to very special cases.

As an alternative, the eddy current method is a viable option for CFRPs, as carbon fibers are electrically conductive. The method is based on the generation of eddy currents within the fibers through an alternating current applied to an inductor probe, which passes over the material a few millimeters from its surface. Sensors detect changes in the resulting magnetic fields, which are then used to map the fiber orientation at a 
particular location~\cite{bardl2016automated, de1992non, lange1994structural,mook2001non, prakash1976eddy, yin2008noncontact}. 
From the resulting measurement data, fiber orientation tensors can be deduced~\cite{bardl2016automated,mook2001non}. Zeng~\cite{zeng2019detection} achieved an accuracy of $0.5^\circ$ for testing fiber orientation or in-plane waviness in CFRP laminates~\cite{romanenko2020materialcharakterisierung}. In current systems, electrically conductive heterogeneities can be visualized. That means, 
local fiber orientation is visible, yet at a spatial resolution too low to render the 
conducting fibers explicitly visible~\cite{Sukhanov_2019}. Commercial systems are also available. One example is the Suragus EddyCus\textsuperscript{{\textregistered}} system~\cite{suragus22}, which combines the eddy current technology with a robotic arm to determine the fiber orientation of carbon composite samples~\cite{heuer2015review}. % and dry textile materials (NCF) [Heuer 2015]. 
The sensors glide over the surface completely without or with slight contact. The required proximity of the sensor to the surface results nevertheless in very slow measurements taking several minutes for only limited areas of a part. 
Often, only region of interest points are scanned, as scanning complete and complex curved geometries remains a challenge. However, using such methods, it seems to be possible to determine localized sub surface information, in particular the local fiber orientation, for laminate depths of up to 4 or 5 layers~\cite{suragus22, suragus23}.

Thermography is a method which takes advantage of anisotropic thermal conductivity for measuring fiber orientations. Heat induced locally in a CFRP sample is conducted preferentially in the direction of the thermally conductive fibers. Thus, the orientation of the fibers can be deduced analytically from the resulting heat map~\cite{aindow1986fibre}.
In advanced systems, several points on the surfaces of CFRP laminates or
parts are heated locally at short time intervals and, subsequently, measured using a thermal imaging camera. This enables analysis of larger surfaces comparatively fast~\cite{karpen1994depth}. However, Fernandes et~al.~\cite{fernandes2013use} report measurement errors of up to $13.5^\circ$ when determining local fiber orientations. Analogous to the eddy current method, spatial resolution and object visibility are therefore also quite limited.

Fiber orientations can also be extracted from images taken by various optical methods, such as classical light microscopy or confocal laser scanning microscopy~\cite{eberhardt2001fibre, hayes10Optical, lee2003measurement}. However, these imaging methods are often restricted to the material's surface or surface-adjacent layers. Examples of commercially available optical systems that are used industrially include but are not limited to DRAPETEST~\cite{textechno22} and ProFactor's FScan technology~\cite{profactor2, profactor1}. The automatic inspection system DrapeWatch~\cite{cikoni22, swinburne22} combines an optical system with eddy current sensors for the selective detection of fiber orientation or internal defects. Yet, these methods are also limited by the size of the devices, required proximity to the scan object, and scan reconstruction time, which can take minutes.
Systems combining 3D laser scanners for capturing surface texture together with high resolution optical cameras as e.g. Hexagon's APODIUS Vision System 3D~\cite{hexagon22} or its online version APODIUS ContInspec~\cite{ccMagazin18} are reported to resolve surface
fiber orientation information well and in real time. They are, however, no longer available.

Döbrich et~al.~\cite{dobrich2023machine} proved that standard optical methods can also be used efficiently. They apply a simple and relatively cheap Microsoft's Azure Kinect RGB D camera 
for mapping the fiber orientation from an optical image onto a 3D mesh of the surface captured by a medium resolution depth sensor on the same device. 
Yet, just like other optical methods, Döbrich's relies on extensive post-processing of the images for edge detection and subsequent calculation of fiber orientation. 
Especially multidirectional C-SMC material requires a resolution high enough to distinguish fibers, which would only be possible with a high number of images. 
The Microsoft Azure Kinect camera used in Döbrich's work is equipped with a 1-MP depth sensor(IR-TOF), a 12-MP RGB camera and a maximum refresh rate of 30 fps.  In October 2023, Microsoft discontinued production of this product. The equipment is a very cheap equivalent to the APODIUS VISION SYSTEM 3D as the depth sensor allows the construction of the 3D surface mesh to which the optical data is mapped. Due to the limited framerate usage is however more for part analysis and not appropriate for online/inline usage.

A further optical method for measuring fiber orientation utilizes the polarizing effects of the observed material/structure. The technology is based on the observation that carbon fibers linearly polarize unpolarized light from natural or artificial light sources by reflection~\cite{freitag2015theoretical}. Fiber orientations can be determined analytically via the measured amplitudes of the input and transmitted or reflected waves, respectively, using the Stokes parameters ~\cite{stokes1851composition} and the set of simple equations defining their relation. 
Microwave radiation yields good results for thin-walled (<\,1\,mm) CFRP specimens like unidirectional prepregs and single dry fabric layers~\cite{urabe1992rotative, urabe1991nondestructive}. 
Application to glass fiber reinforced polymers (GFRP) is also possible, but rather inaccurate, due to similar reflection and diffraction properties of the glass fibers and the polymer matrix.

The technologies discussed so far and their applicability to fiber orientation measurement of CFRPs are summarized in Table~\ref{tab:imaging}.
\begin{table}[ht]
\begin{flushleft}
\caption{Overview of selected non-destructive testing methods suitable for fiber orientation measurement in glass or carbon fiber-reinforced composites~\cite{de1992non, gulhane2023advance, schoberl2016measuring, wang2020non}.\label{tab:imaging}}

\begin{tabular}{ | l | l | l|l| c| c |}
\hline
\multicolumn{2}{|l|}{Test Procedure} & \multicolumn{2}{c|}{Materials} & \multicolumn{2}{c|}{Possible precision of fiber orientation measurement}\\
\hline
Principle & Method & CFRP & GFRP & \phantom{texttextt} 2D \phantom{texttextt}&\phantom{texttextt} 3D \phantom{texttextt}\\
\hline \hline
\multirow{2}{*}{X-ray contrast} & Radiography & yes & yes & medium & qualitative only\\
\cline{2-6}
 & Computed tomography & yes & yes & high & high \\
\hline
Heat conduction & Thermography & yes & no & medium & qualitative only\\
\hline
Electrical conduction & Eddy current & yes & no &medium & qualitative only\\
\hline
\multirow{3}{*}{Optical} & Optical microscopy & yes & yes & high & impossible \\
\cline{2-6}
 & Confocal laser scanning & yes & yes & medium & medium \\
\cline{2-6}
 & Polarization imaging & yes & no & high & qualitative only\\
\hline

\end{tabular}
\end{flushleft}
\end{table}

\subsection{Validating Polarization Imaging for Orientation Analysis}

Looking for a machine vision system with realistic potential for real-time process control and digitalization of carbon fiber based polymer composite materials, we opt for an image processing system on the basis of compact, easy-to-use real-time polarization imaging. The measured angle of linear polarization (AOLP) reflected by the carbon fibers is directly related to the fiber orientation. 
AOLP measurement is possible thanks to a special on-pixel polarization filter sensor, patented by the Fraunhofer Institute for Integrated Circuits
~\cite{atkinson2018high, ernst2014messung, ernst2016measurement, schoberl2016measuring} and later commercialized by Sony in 2018. Today, these sensors can be found in a range of commercially available, compact USB polarization cameras costing between €1\,000 and €3\,000.

We have applied this type of polarized machine vision for the quality control of C-SMCs~\cite{duhovic2022digitizing, schommer2023polarization}. Shen et al.~\cite{shen2023} also employed the technology to validate decoupled orientations of polymer chains and carbon fibers.
The method has been validated with a promising accuracy of around 1$^\circ$~\cite{atkinson2018high, atkinson2021precision, chiominto2024using} but for uni- and bidirectional material only and not for more general use cases as realistic multi-directional C-SMC materials. 
Moreover, polarized machine vision is in principle limited to capturing fiber orientations only on the material's surface. It has however never been explored, to what extent this information correlates with fiber orientations in deeper layers of the material.

To shed more light on these two questions, we imaged press rheometry test specimens of 
a selection of customary C-SMC materials using a commercially available, compact USB polarization camera~\cite{ernst2016measurement} to generalize the validation study ~\cite{atkinson2018high} to a realistic industrial setting. 
Moreover, we validate the polarization images against CT images of the same specimens as the latter also provide images of interior material layers. 
Based on the CT images, we show that the fiber orientation distribution in surface layers of our specimen is representative for the interior layers. 
Finally, we discuss the limitations of the polarized machine vision system when applied to C-SMC and provide an outlook for future research.

\section{Material and Methods}

In this section, we describe the C-SMC materials used, the parameters of the 
press rheometry test and the two applied imaging methods. 
We recall the 2D orientation analysis by maximal response of anisotropic Gaussian filters and discuss how to quantify the deviation of 2D orientation maps.

\subsection{Material \& Sample Preparation}
\label{sec:material}

One aim of this study was to validate the polarized machine vision system such that it can be used for all commercially available C-SMC materials, solely relying on the data sheet provided by the supplier, and without further information regarding the exact formulation of the material. In order to create this situation, a selection of commercially available C-SMC materials have been chosen.

Polynt Composites' SMCarbon\textsuperscript \textregistered  24 CF50-3K~\cite{SMCarbon3k, SMCarbon3kt} is a vinyl ester based C-SMC, which is regularly used in the automotive industry\,\cite{CompoundsCarbonFiber}. This material has a fiber weight content of 50\,\% and a long fiber reinforcement consisting of 3K (approximately 3\,000 individual fibers) yarns with a length of 25\,mm. 
The yarns are spread in layers to form a 2D reinforcement structure. The semi-finished product is supplied in rolls with a usable width of 500\,mm and a customizable basis areal weight. In order to investigate the effects of different yarn sizes, we also analyzed SMCarbon\textsuperscript \textregistered  24 CF50-12K~\cite{SMCarbon12k, SMCarbon12kt} from the same product range. The only difference between the two materials is the yarn size (12K $\approx$ 12\,000 individual fibers) with an identical fiber weight content of 50\,\%. For this study, both materials were ordered with a basis areal weight of 1\,400\,g/m$^2$, for better comparability of the analyzed materials.

LyondellBasell's Quantum AMC\textsuperscript \textregistered 85593~\cite{AMC85593} and Quantum AMC\textsuperscript \textregistered  85590~\cite{AMC85590} were chosen for their comparability to the Polynt materials. According to the manufacturer, these two materials are based on the formulations of AMC\textsuperscript \textregistered  8593~\cite{AMC8593t} and AMC\textsuperscript \textregistered  8590~\cite{AMC8590t}, (available in the US) but are produced in Europe. Like the Polynt materials, these C-SMCs have a vinyl ester matrix and polyacrylonitrile-based 3K or 12K yarns. 
The fiber content is also specified as 50\,\% by weight. As this manufacturer also defines the basis areal weight of the semi-finished product customer-specifically, there is no information on this on the data sheet. A basis areal weight of 1\,400\,g/cm$^2$ was ordered for comparability with the Polynt products. A summary of all materials used in the study is given in Table~\ref{tab:ManufacturerInfo}.

\begin{table}[ht]
\caption{Summary of material properties of the analyzed commercially available C-SMC semi-finished products.\label{tab:ManufacturerInfo}}
\begin{center}
\begin{tabular}{ |l || c | c | c | c |}
\hline
Manufacturer & \multicolumn{2}{c|}{Polynt Composites} & \multicolumn{2}{c|}{LyondellBasell}\\
\hline \hline
Description & SMCarbon\textsuperscript \textregistered 24  & SMCarbon\textsuperscript \textregistered 24  &
AMC\textsuperscript \textregistered  & AMC\textsuperscript \textregistered \\
&CF50-3K~\cite{SMCarbon3kt}&CF50-12K~\cite{SMCarbon12kt}
&85593~\cite{AMC8593t} & 85590~\cite{AMC8590t}\\
 \hline
Roving weight percentage [\%] & 50 & 50 & 50 & 53\\
\hline
Roving length [mm] & 25.4 & 25.4 & 25.4 & 25.4\\
\hline
Roving type & 3K & 12K & 3K & 12K\\
\hline
Areal weight [g/m\textsuperscript{2}] & 1\,400 & 1\,400 & 1\,400 & 1\,400\\
\hline
Density (cured) [g/cm\textsuperscript{3}] & 1.43 & 1.40 & 1.47 & 1.48\\
\hline

\end{tabular}
\end{center}
\end{table}
The tests were conducted under the "constant mass" configuration. In these characterization experiments, specimens 100\,mm\,$\times$\,100\,mm in size are pressed between two parallel flat plates to various thicknesses ranging from initial approximately 15\,mm to 2-3\,mm, typical for C-SMC parts manufactured for a variety of industrial applications. These types of press rheometry tests are termed "constant mass" as the quantity of material between the plates remains constant. The shape of the outer boundary of the pressed specimens indicates the degree of material isotropy or anisotropy. The specimen starts out square and then becomes more circular while being pressed and the material flows outwards radially. 
Our samples were pressed at mold closing speeds of 0.5\,mm/s and 3.0\,mm/s, respectively (see Table~\ref{tab:rheometry} for details) to a final thickness of 3\,mm, which correspond to typical industrial processing conditions.

Differences in material behavior during compression molding can result from differences in the matrix formulation, the types of carbon fibers used, their sizing or surface treatments as well as influences from the production process of the semi-finished product itself. Such detailed information regarding material production is generally not known to the user of the semi-finished products. 
A smaller subset of eight press rheometry test specimens from a larger test series of 108 specimens 
(4 materials, 3 testing speeds, 3 short-shot distances, and 3 repeats, see also \cite{romanenko2022process-simulation, schommer2019material}) was taken for the validation study as summarized in Table~\ref{tab:rheometry}.

\begin{table}[ht]
\caption{Summary of analyzed press rheometry test specimen series.\label{tab:rheometry}}
\begin{center}
\begin{tabular}{ |l || c | c | c | c | }
\hline
Specimen & Material & Velocity [mm/s] & Short shot distance [mm] & Roving size\\
\hline \hline
\phantom{1}19 & SMCarbon\textsuperscript \textregistered 24 CF50-12K & 0.5 & 3 & 12K\\
\hline
\phantom{1}33 & SMCarbon\textsuperscript \textregistered 24 CF50-12K & 3.0 & 3 & 12K\\
\hline
\phantom{1}53 & SMCarbon\textsuperscript \textregistered 24 CF50-3K & 0.5 & 3 & 3K\\
\hline
\phantom{1}66 & SMCarbon\textsuperscript \textregistered 24 CF50-3K & 3.0 & 3 & 3K\\
\hline
\phantom{1}86 & AMC\textsuperscript \textregistered 85590 & 0.5 & 3 & 12K\\
\hline
\phantom{1}97 & AMC\textsuperscript \textregistered 85590 & 3.0 & 3 & 12K\\
\hline
120 & AMC\textsuperscript \textregistered 85593 & 0.5 & 3 & 3K\\
\hline
131 & AMC\textsuperscript \textregistered 85593 & 3.0 & 3 & 3K\\
\hline

\end{tabular}
\end{center}
\end{table}

\subsection{Polarization Imaging and Computed Tomography}
\label{sec:imaging}

We imaged each specimen with the two modalities -- a polarization camera and an X-ray microscope. First, the front and back of the specimen were scanned using the polarization camera VCXU-50MP by Baumer GmbH with an exposure time of 20\,ms and a pixel spacing of approximately 0.21\,mm.
\begin{figure}[ht]%
\centering
\includegraphics[width=\textwidth]{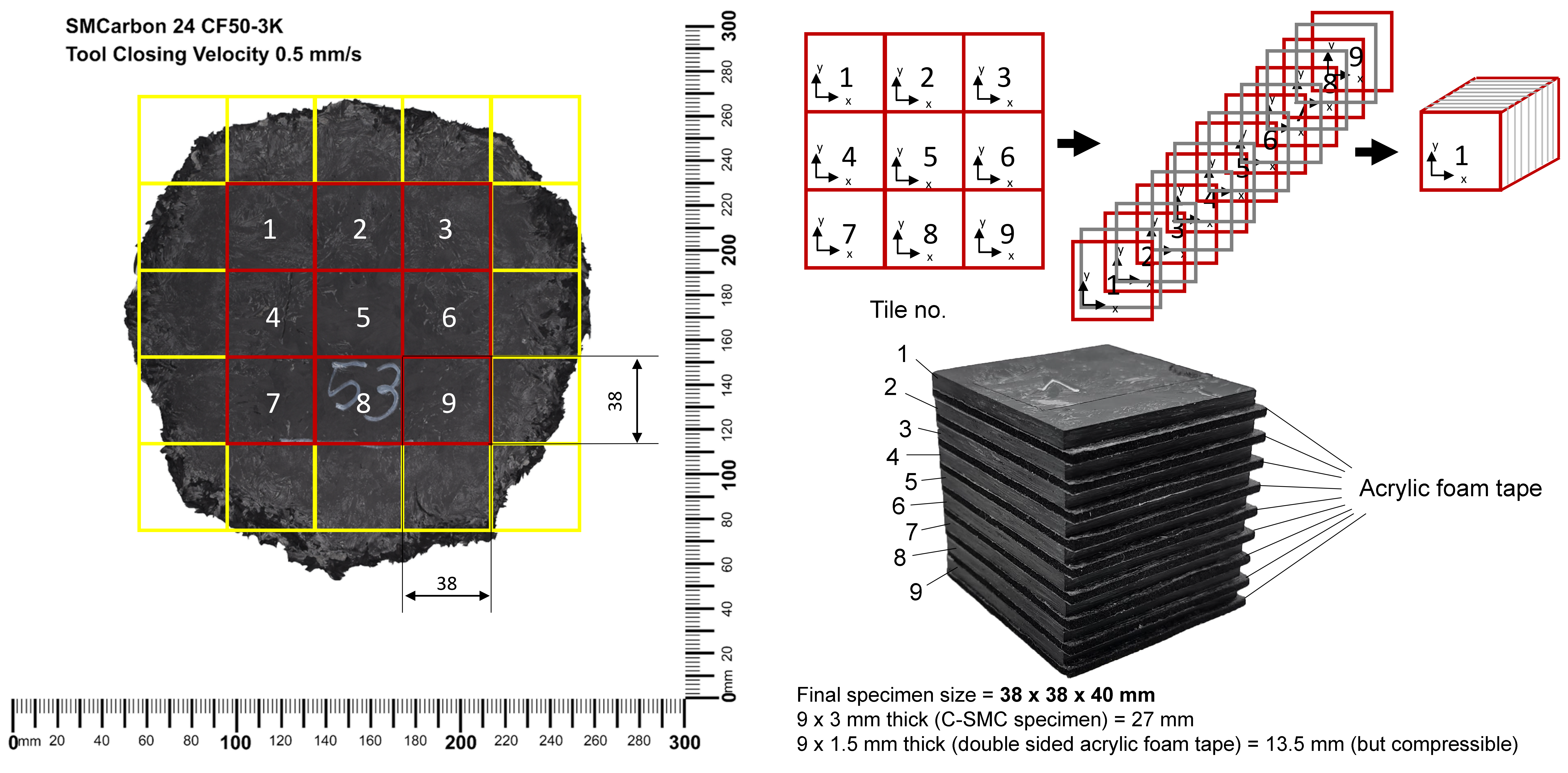}
\caption{3D CT scan sample preparation of a full press rheometer specimen from the "constant mass" short-shot configuration.\label{fig:samplePreparation}}
\end{figure}
The 3D CT scans were carried out on an Xradia 520 Versa, manufactured by Carl Zeiss Mikroskopie GmbH. Scanning carbon fibers with an X-ray microscope is generally challenging because they provide little contrast to the resin. Moreover, both high resolution and a large field of view are necessary to reliably validate the polarization camera images. Hence, the Xradia 520 Versa is well-suited for resolving and visualizing carbon fibers due to its high resolution and ability to provide phase contrast. An illustrative example of one of the eight specimens taken from the press rheometry tests is shown in Fig.~\ref{fig:samplePreparation}. The scan quality could have only been improved using synchrotron radiation, which was outside the scope of this paper.

Given the substantial difference between the in-plane dimensions and the thickness of the specimen, scanning the specimen as is would result in considerable variation in the distances the X-ray beams have to pass through the material, leading to a significant loss of scan quality. 
To address this issue, the sample was divided into 38\,mm $\times$ 38\,mm cut-outs using a Mutronic DIADISC 6200 precision saw with a cutting blade thickness of 2.5\,mm. 
These cut-outs were then rearranged to achieve a cuboid shape of the specimen as shown in Fig.~\ref{fig:samplePreparation}. Smaller cut-outs would have yielded a higher resolution with higher contrast. However, this would also have resulted in either a smaller area to be scanned, or in more cutting borders, which are a considerable source of error.
Due to their undefined shape, we omitted the cut-outs at the edge areas of the specimen and stacked only the center areas 1-9 into a nearly cubic cuboid. In addition, attention was paid to the orientation of the tiles during stacking so that they could be reconstructed properly into the specimen’s original shape. In order to facilitate the identification of individual tiles in the subsequent CT image, they were separated by double-sided acrylic foam adhesive tape.

\begin{figure}[ht]%
\centering
\includegraphics[width=\textwidth]{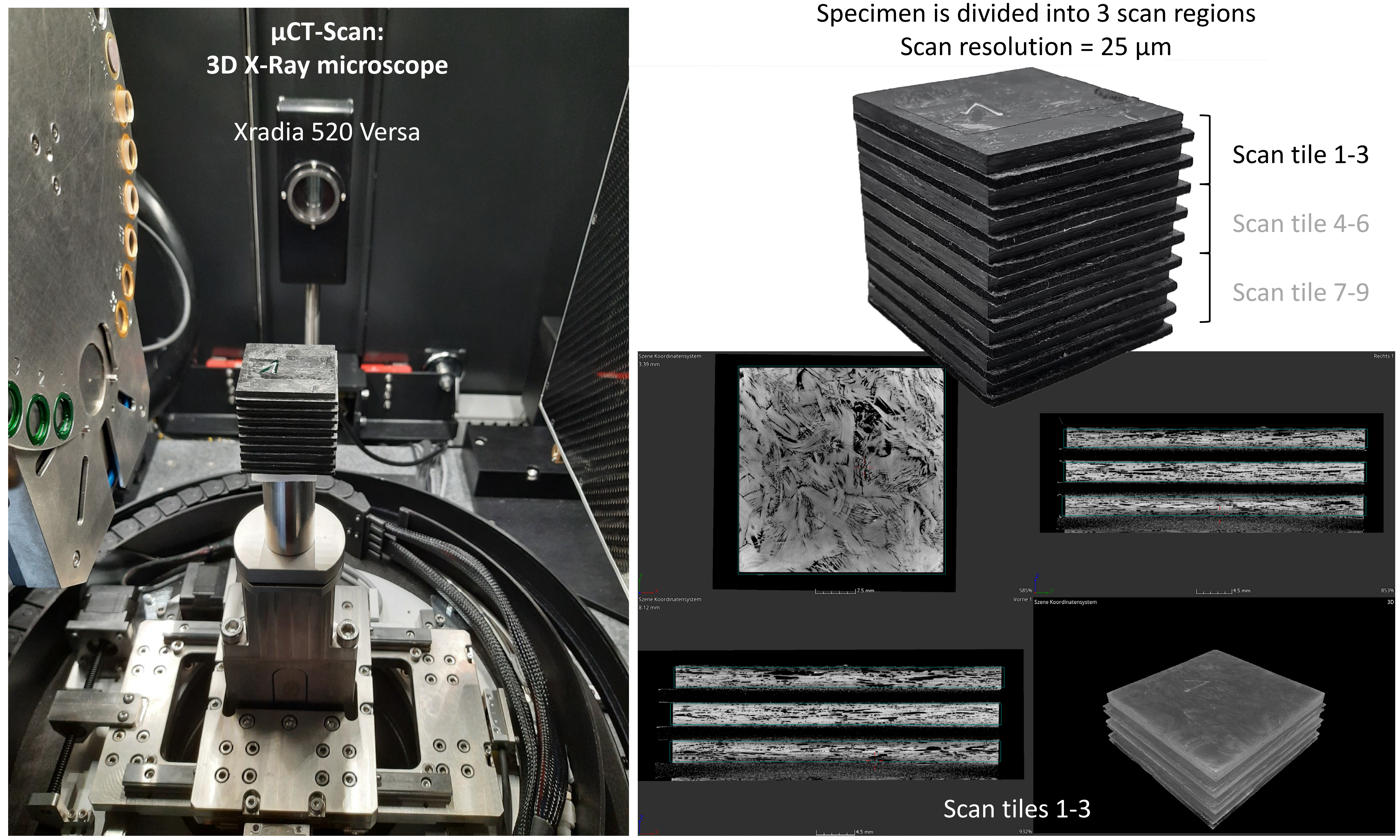}
\caption{Left: Positioning of the C-SMC specimen in the scanning chamber of the Xradia 520 Versa 3D X-ray microscope. Right:  \textmu CT scan of the stacked tiles 1-3 of specimen 53, see Table~\ref{tab:rheometry}.\label{fig:positionScanning}}
\end{figure}

To ensure optimal scan quality for the entire specimen, the nine tiles were scanned separately in groups of three, as shown in Fig.~\ref{fig:positionScanning}. Each scan covered a field of view measuring 76\,mm $\times$ 48\,mm, with nominal resolution (pixel size) 25\,\textmu m. Operating at 5\,W power and a voltage of 60\,keV, each scan involves 3\,201\,projections. During a brief exposure time of 2\,sec per projection, a total of 20 individual images were captured at 0.1\,s intervals and were averaged subsequently. Due to the phase contrast exploited here, matrix material and carbon yield a good gray value contrast for specimen 53 shown in Fig.~\ref{fig:positionScanning}. However, there is nearly no contrast or edge information within the fiber rovings. Moreover, the contrast is significantly lower for other specimens, see e.g. Fig.~\ref{fig:CTundMR}. Yet, we can estimate the fiber directions accurately from the CT scans using the maximal response of anisotropic Gaussian filters~\cite{robbFiberOrientationEstimation2007}, as we will discuss in the following.

\subsection{Fiber Orientations from the CT Image by Maximal Response of Anisotropic Gaussian Filters}
\label{sec:oriEstimation}

Estimating the fiber orientation of C-SMC in CT images is challenging due to the low contrast between fibers and matrix. 
Therefore, we estimate the fiber orientations from the CT images using a method that is particularly well suited to low-contrast images, namely the maximal response of anisotropic Gaussian filters \cite{robbFiberOrientationEstimation2007} (MR method), as suggested by Schladitz et~al.~\cite{schladitzNondestructiveCharacterizationFiber2017} for glass fiber reinforced sheet molding compounds.

Anisotropic Gaussian filters can model the orientation of fibers well due to their similarly elongated shape. Hence, they are used as pattern-matchers: The response of an anisotropic Gaussian filter to a patch of a fiber image is maximal when aligning it with the fiber. Then, its orientation coincides with the fiber's orientation (up to a discretization error). Following this reasoning, the MR method applies anisotropic Gaussian filters over a set of sampled orientations to the whole image. For each pixel, the maximal response and the corresponding orientation are returned.

Due to the production process of SMC, the fibers lie within a plane. Hence, estimating the fiber orientations based on 2D image slices is sufficient.
For that purpose, we preprocess the CT images using the software MAVI \cite{mavi}: We crop the images to obtain one subvolume per tile. We rotate each such image such that the tile surface is aligned with the image's x-y-plane. 
The fiber orientations are then computed for each x-y-slice of the rotated images. 

We process the image slices using ToolIP \cite{toolip}. First, we equalize gray values by applying a mean filter with mask size~49 and subtracting the result from the original image.
Next, we estimate the fiber orientation by applying the MR method as implemented by Keilmann et al.
\cite{Keilmann_Godehardt_Moghiseh_Redenbach_Schladitz_2024}, where we employ cubic interpolation. 
We use the variances $\sigma_1~=~25.0, \sigma_2~=~2.0$ for the Gaussian filters 
since these are the most accurate variances determined in 
\cite{Keilmann_Godehardt_Moghiseh_Redenbach_Schladitz_2024} that can still imitate the shape of a 
fiber. We restrict the possible orientations to $\theta = 0^\circ, 1^\circ, ..., 179^\circ$.  Note 
that we omit $\theta = 180^\circ, 181^\circ, ..., 359^\circ$ as the choice between $\theta$ and 
$\theta + 180^\circ$ is ambiguous for fibers.

In order to exclude orientation measurements that are not well-defined, we threshold the response with Niblack's method \cite{niblack1985introduction} using a window size of $w = 4 \sigma_2$ following Schladitz et~al.~\cite{schladitzNondestructiveCharacterizationFiber2017}, and the threshold $0.6$. In addition, we exclude components of the mask that are very small, i.e. we erode the mask with a square of size $3\times 3$ pixels, then exclude connected components with a size smaller than 100 pixels, and dilate the mask with the same square again.

Like Schladitz et~al.~\cite{schladitzNondestructiveCharacterizationFiber2017}, we exclude all pixels that are $2\sigma_1$ close to the image border.
Finally, we stitch the results of the top layers of the tiles back together, padding areas that were excluded or not scanned (see Fig.~\ref{fig:samplePreparation}).

\subsection{Similarity Measures for 2D Orientation Data}
\label{sec:error+registration}
To validate the orientation measurement via polarization imaging using the MR results, we first need to align the images from both modalities. This means that pixels showing the same point in the specimen have to be identified in both images. 
This procedure is called registration~\cite{gonzalez2018DigitalImageProcessing}.
 First, we explain how the polarization and CT images are related. Then, we introduce the needed measure for image similarity, which we later also use for validating the polarization images. Subsequently, we formalize our registration process, and finally, introduce deviation categories for validation.

 \begin{figure}[h]
    \centering
    \begin{subfigure}{0.32\textwidth}
        \centering
        \includegraphics[width=\textwidth]{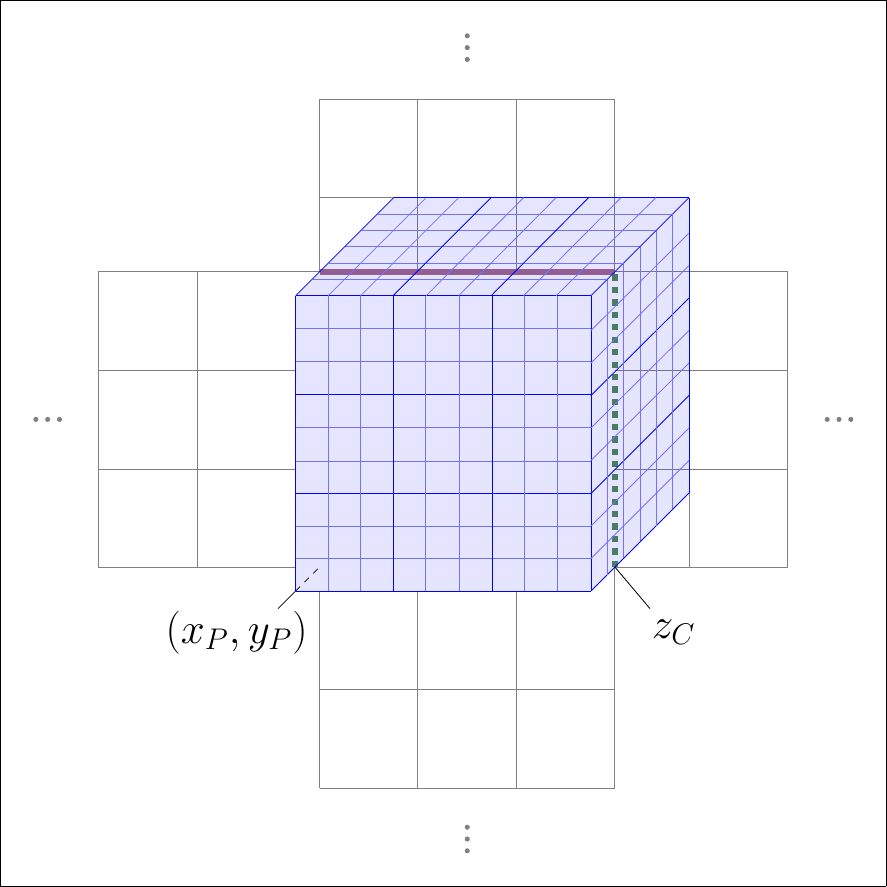}
        \caption{Schematic visualization of the full registration process.}
    \end{subfigure}
    \hfill
    \begin{subfigure}{0.32\textwidth}
        \centering
        \includegraphics[width=\textwidth]{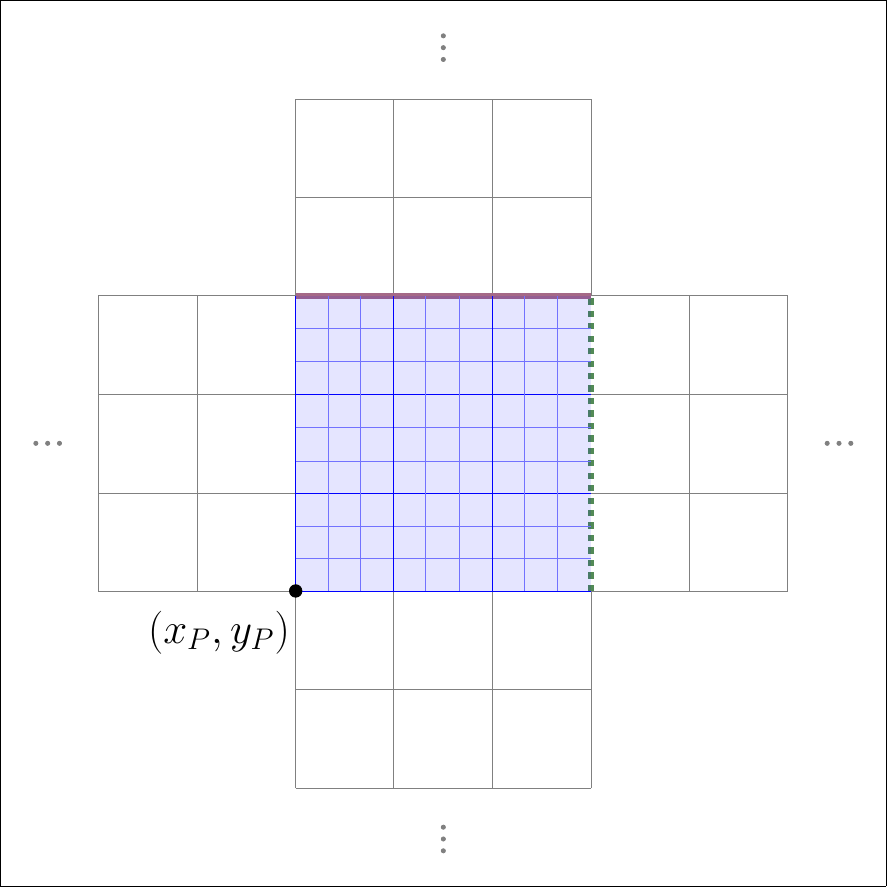}
        \caption{Schematic visualization of registration w.r.t. the x-y-position.}
    \end{subfigure}
    \hfill
    \begin{subfigure}{0.32\textwidth}
        \centering
        \includegraphics[width=0.21\textwidth]{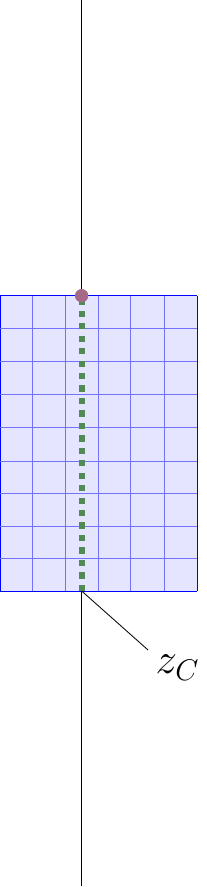}
        \caption{Schematic visualization of registration w.r.t. the z-slice.}
    \end{subfigure}
    \caption{Schematic visualization of the registration process. The polarization image is depicted in black-white, whereas the CT image is depicted in shades of blue. For reference between depictions, one shared image edge is depicted in red and another one in dashed green.}
    \label{fig:registration}
\end{figure}

Fig.~\ref{fig:registration} illustrates the registration process for our CT and polarization images: The polarization image covers only the top or bottom surface of the specimen, yet its whole extent. It has a size of $1\,232 \times 1\,028$ pixels, and its pixel indices are denoted as $x_p, y_p$. The CT images, on the other hand, cover the whole depth of the specimen, but only for a specific tile of the specimen. They have a size of $1\,520 \times 1\,520 \times z$ voxels where $z$ varies due to the varying thickness of the specimen. 
However, with a voxel edge length of 25\,\textmu m, the CT images have a far higher resolution than the polarization images whose pixels have an edge length of 210\,\textmu m. For comparability, we downscale the x-y-slices by a factor of 8.4 with the new pixels being assigned the mean of pixel values within the mask. 

\begin{figure}
    \center
    \begin{tikzpicture}
        \path (20:2cm) coordinate (P0);
        \path (200:2cm) coordinate (P0e);
        \path (160:2cm) coordinate (P1);
        \path (340:2cm) coordinate (P1e);
        \path (0:2cm) coordinate (Pm);
        \path (180:2cm) coordinate (Pme);
        \path (90:2cm) coordinate (Pam);
        \path (270:2cm) coordinate (Pame);
        \draw (0,0) circle(2cm);
        
        \draw[-,  line width=2pt] (P0e) -- (P0) node[black,pos=1.1] {$20^\circ$}; 
        \node[draw,circle,inner sep=1.5pt,fill] at (P0) {};
        \draw[-,  line width=2pt] (P1e) -- (P1)  node[black,pos=1.1] {$160^\circ$};
        \node[draw,circle,inner sep=1.5pt,fill] at (P1) {};
        \draw[-,  line width=2pt, dashed, green] (Pme) -- (Pm) node[black,pos=1.1] {$0^\circ$}; 
        \node[draw,circle,inner sep=2pt,fill, green] at (0:2cm) {} ;

        \draw[-,  line width=2pt, dashed, red] (Pame) -- (Pam) node[black,pos=1.1] {$90^\circ$}; 
        \node[draw,circle,inner sep=2pt,fill, red] at (90:2cm) {} ;
    \end{tikzpicture}
    \caption{Illustration of pitfalls when calculating the mean of angular values.\label{fig:angleMean}}
\end{figure}
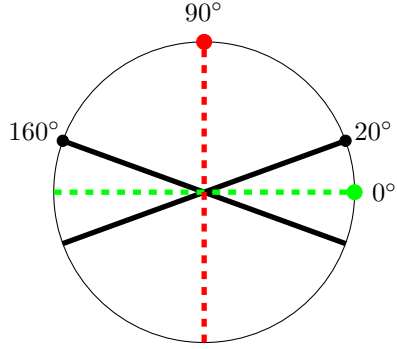

Note that angles cannot be averaged in a straightforward way~\cite{mardia2000DirectionalStatistics}: For example, the arithmetic mean of $20^\circ$ and $160^\circ$ is $90^\circ$, while the accurate mean should yield $0^\circ$ for axial data, see Fig.~\ref{fig:angleMean}. Therefore, we first calculate the center of mass $(\bar{C}, \bar{S})$ as
\begin{equation*}
    \bar{C} = \frac{1}{n}\sum_{j = 1}^n \cos 2\theta_j,~~~~ \bar{S} = \frac{1}{n}\sum_{j = 1}^n \sin 2\theta_j
\end{equation*}
for angles $\theta_j, j = 1, ..., n \in \mathbb{N}$. Then, the mean angle is given by
\begin{flalign*}
    \bar{\theta} =\begin{cases}
			\frac{1}{2}\arctan(\bar{S}/\bar{C}), & \text{if $\bar{C} \leq 0$}\\
            \frac{1}{2}\arctan(\bar{S}/\bar{C}) + 180^\circ, & \text{otherwise,}
		 \end{cases}
\end{flalign*}
see~\cite{mardia2000DirectionalStatistics}. This yields image slices of size $172 \times 172$ pixels. We denote the voxel indices of the downscaled image slices as $x_{c}, y_{c}, z_c$.

To register the CT and the polarization images, we need to align the CT image correctly with the corresponding position ($x_P, y_P$) in the polarization image on the one hand. On the other hand, we need to find the x-y-slice $z_C$ in the CT image which depicts the surface. For both the registration process and the validation later on, image similarity has to be quantified. We base the similarity measure on the mean absolute error (MAE). 

To take the periodicity of angles into consideration~\cite{mardia2000DirectionalStatistics}, we appropriately adapt the usual formulas of the MAE to the mean angular deviance (MAD) as follows: We calculate the deviation map between the polarization image $p$ and the slice-wise MR result $c$ based on the CT scan as
\begin{flalign*}
    m(x_c, x_p, y_c, y_p, z_c) =\begin{cases}
			\phantom{ 180^\circ - } ~|p(x_p, y_p) - c(x_c, y_c, z_c)|, & \text{if $|p(x_p, y_p) - c(x_c, y_c, z_c)| \leq 90^\circ$}\\
            180^\circ - |p(x_p, y_p) - c(x_c, y_c, z_c)|, & \text{otherwise.}
		 \end{cases}
\end{flalign*}
To calculate the MAD, we take the mean of the deviation map $m$ over every pixel within the mask $M$ when placing the mask at position $(x_P, y_P)$:
\begin{equation*}
    \text{MAD}(x_P, y_P, n_x, n_y, z_c) = \frac{1}{n_x n_y}\sum_{x_c = 0, y_c = 0}^{n_x, n_y} \mathbb{1}_{(x_c, y_c) \in M}m(x_c, x_c+x_P, y_c, y_c + y_P, z_c)
\end{equation*}
for a CT image of size $n_x \times n_y$, so in our case $n_x, n_y = 172$. Note that $x_P < 1\,232 - n_x, y_P < 1\,028 - n_y$ to ensure valid pixel indices of the polarization image.

We first register the downscaled MR result and the polarization image visually for each tile by overlaying the images and moving them in relation to each other, until they seem well aligned. This gives us the indices $x_P', y_P', z_C'$. Next, we precisely register them brute-force by minimizing the MAD. More specifically, we calculate
\begin{equation*}
    (x_P'', y_P'', z_C'') = \underset{\substack{x_P \in [x_P' - 25, x_P' + 25],\\ y_P \in [y_P' - 25, y_P' + 25],\\ z_c \in [0, 15]}}{\text{argmin}} \text{MAD}(x_P, y_P, n_x, n_y, z_c).
\end{equation*}

Subsequently, for each pixel $(x_c,y_c)$, we pick the slice $z_C'''$ of the MR result that minimizes the MAD w.r.t. the registered polarization image:
\begin{equation*}
    z_C''' = \underset{\substack{ z_c \in [0, 15]}}{\text{argmin}} m(x_c, x_c + x_P'', y_c, y_c + y_P'', z_c)
\end{equation*}
This provides a way of mimicking possible shine-through effects in polarization imaging.

For each specimen, we report the MAD after registration, i.e. $\text{MAD}(x_P'', y_P'', n_x, n_y, z_C''')$. Moreover, we combine both the results for the upper and lower surface of the specimen. Note that for the lower surface, the polarization image has to be mirrored regarding its position and angular pixel value, and for the CT images, the lowest slices must be used. Additionally, we present the area fraction of the following deviation categories:
\begin{itemize}
    \item Very low: $[\phantom{1}0^\circ, \phantom{1}5^\circ)$
    \item Low: ~~~~~~~~$[\phantom{1}5^\circ, 10^\circ)$
    \item Fair: ~~~~~~~~~$[10^\circ, 20^\circ)$
    \item High: ~~~~~~~$[20^\circ, 90^\circ]$
\end{itemize} 

\subsection{Comparison of Layerwise Orientation Distributions}
\label{sec:kernelDensity}

The present SMC material consists of multiple fabric layers stacked on top of each other. Yet, with a polarization camera, only surface images can be taken, whereas CT also provides images of internal layers. Hence, we aim to determine how much information on the orientation distribution for the whole specimen is lost when using only surface images. For this, we compare the histograms of interior image slices with the surface image slices as registered in Section~\ref{sec:error+registration}. We select the central image slice for each material layer for LyondellBasell's materials, yielding 4 equidistant image slices. As the Polynt material has 8 material layers, we analogously select the central image slice for every other layer to ensure comparability.

Instead of the raw histograms, we present the kernel density estimates using Sheather \& Jones' method~\cite{sheather91} with a factor of 2 to estimate the bandwidth and the Gaussian kernel for smoothing. For periodicity treatment, we extend the histogram to $(-180^\circ, 360^\circ)$ by replicating the original values.

\section{Results}
\label{sec:results}
\begin{figure}[!h]
\centering
\begin{subfigure}{0.49\textwidth}
    \centering
    \includegraphics[height=.95\linewidth]{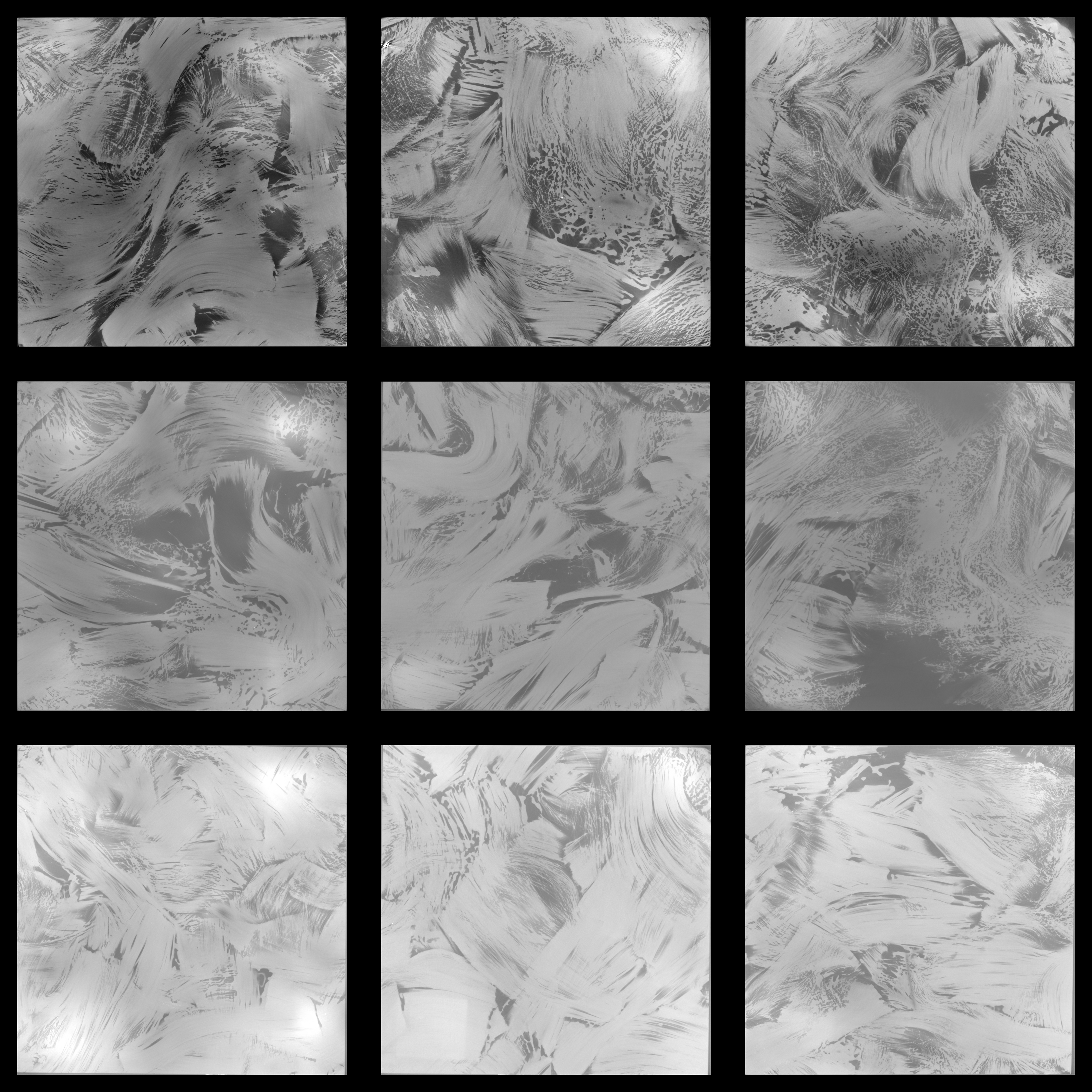}
    \caption{Spread slices of the \textmu CT scans stitched together.\\~}
\end{subfigure}
\begin{subfigure}{0.49\textwidth}
    \centering
    \includegraphics[height=.95\linewidth]{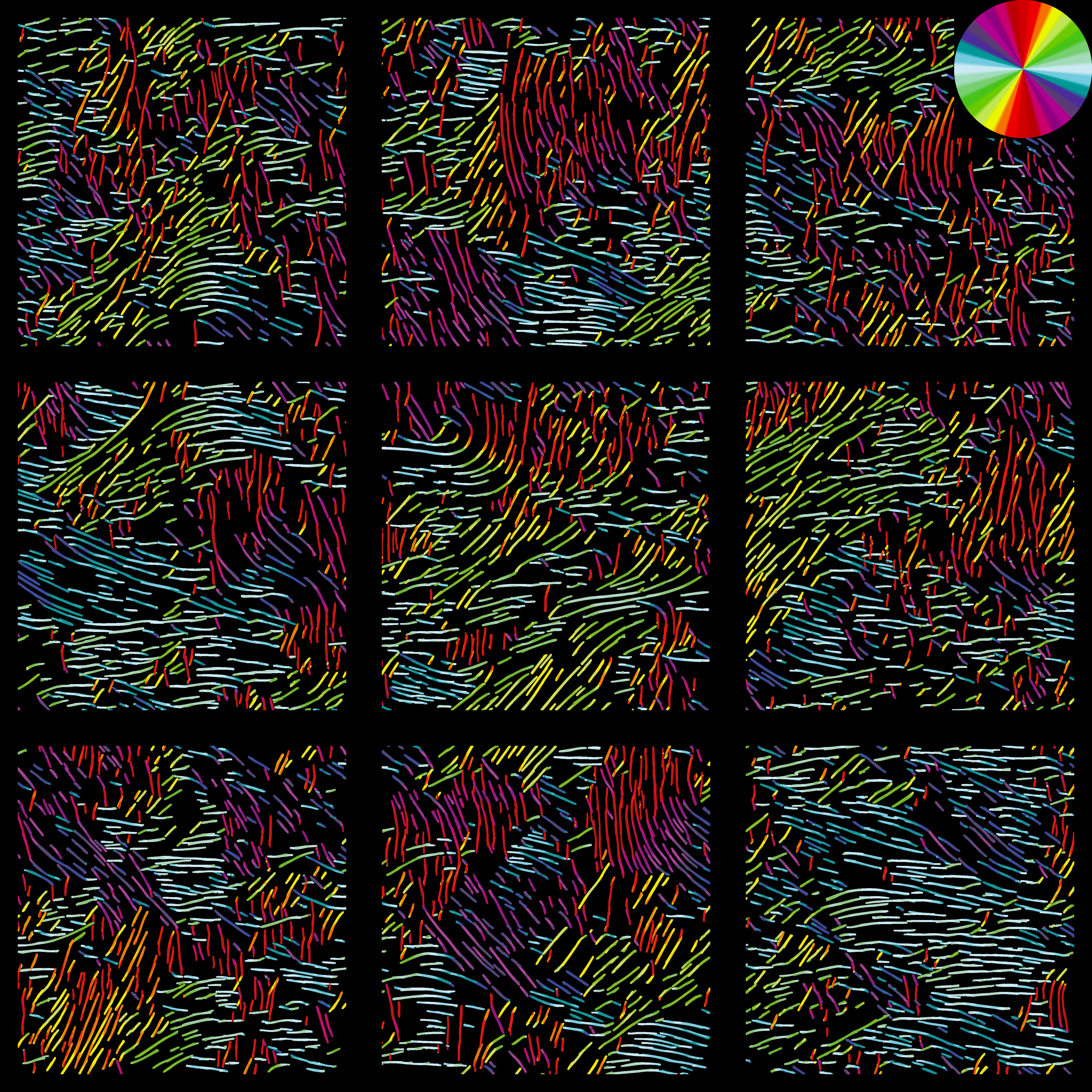}
    \caption{Fiber orientations after applying the MR method to the \textmu CT scans (dilated for better visibility).}
\end{subfigure}
\caption{Surface layer of specimen 19 (\textmu CT scans).}
\label{fig:CTundMR}
\end{figure}

In order to validate the polarization images on \textmu CT scans, we first calculated the local fiber orientations from the \textmu CT images using the MR method. The \textmu CT images are inevitably of low resolution and low contrast since we aimed at a large field of view. Nevertheless, the MR method yields fairly robust results, see Fig.~\ref{fig:CTundMR}.

\begin{figure}[!h]
\centering
\begin{subfigure}{0.32\textwidth}
    \centering
    \includegraphics[width=.95\linewidth]{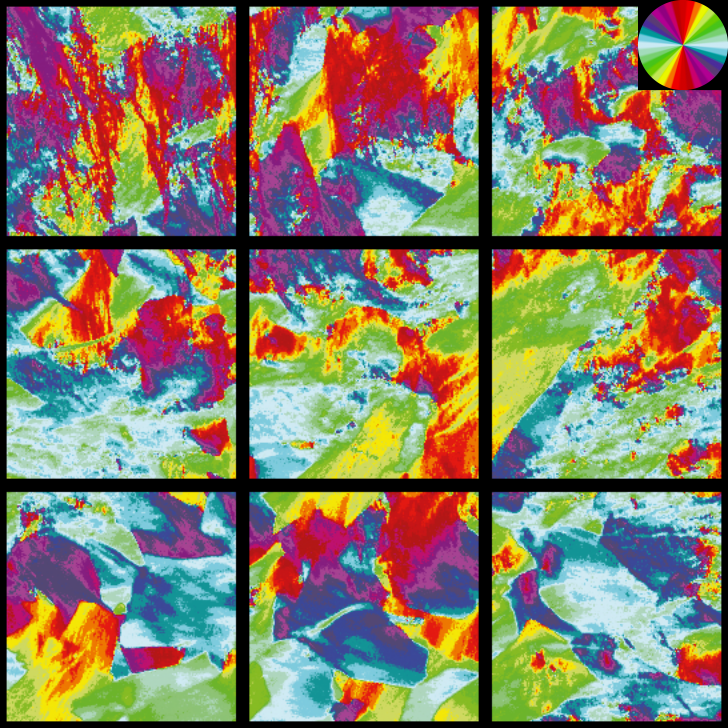}
    \caption{Results of the polarization camera after registration.}
    \label{fig:19t-POLKA}
\end{subfigure}
\begin{subfigure}{.32\textwidth}
    \centering
    \includegraphics[width=.95\textwidth]{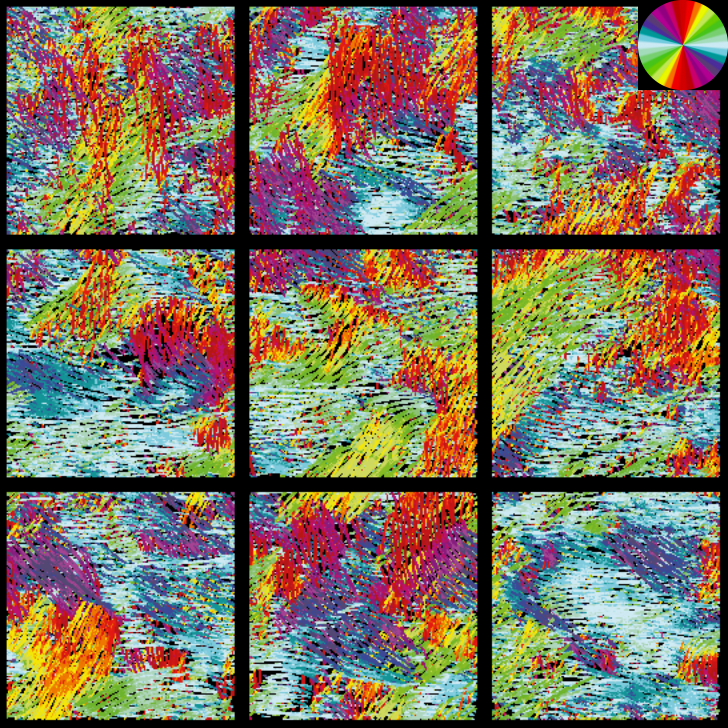}
    \caption{MR results (downsampled \textmu CT) after registration.}
    \label{fig:19t-stitched}
\end{subfigure}
\begin{subfigure}{0.32\textwidth}
    \centering
    \includegraphics[width=.95\linewidth]{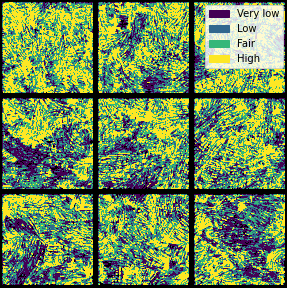}
    \caption{Deviation plot between polarization images and MR results.}
    \label{fig:19t-error}
\end{subfigure}
\caption{Results of the top layer from specimen 19 for both imaging modalities.}
\label{fig:19t}
\end{figure}

Next, we registered the downsampled MR results and the polarization images. As can be seen in Fig.~\ref{fig:19t}, the images seem well-matched and appear remarkably similar. Note though that the color wheel covers $10^\circ$ per color. A substantial area features deviations above $20^\circ$  as the plot in Fig.~\ref{fig:19t-error} reveals.
We calculated the exact deviation areas and the MAD for all specimens and listed them in Table~\ref{tab:MAEandError}. The MAD was consistently around $20^\circ$ and well over half the area was estimated with a deviation below $20^\circ$.

\begin{table}[!h]
\caption{Mean angular deviation (MAD) and area fractions of deviation categories for the surface layers of each specimen.}
\label{tab:MAEandError}
\begin{center}
\begin{tabular}{ |l || c | c | c | c| c | }
\hline
Specimen & MAD & Very low&Low & Fair & High\\
\hline \hline
\phantom{1}19 & $18.33^\circ$ & 0.32 & 0.17 & 0.19 & 0.32\\
\phantom{1}33 & $21.28^\circ$ & 0.26 & 0.15 & 0.20 & 0.40\\
\phantom{1}53 & $21.13^\circ$ & 0.27 & 0.16 & 0.19 & 0.39\\
\phantom{1}66 & $23.02^\circ$ & 0.23 & 0.15 & 0.19 & 0.43\\
\phantom{1}86 & $23.74^\circ$ & 0.23 & 0.15 & 0.19 & 0.44\\
\phantom{1}97 & $21.68^\circ$ & 0.25 & 0.15 & 0.19 & 0.40\\
120 & $26.60^\circ$ & 0.17 & 0.12 & 0.19 & 0.52 \\
131 & $25.95^\circ$ & 0.18 & 0.13 & 0.19 & 0.50\\
\hline

\end{tabular}
\end{center}
\end{table}

\begin{figure}[!ht]%
\centering
\includegraphics[width=\textwidth]{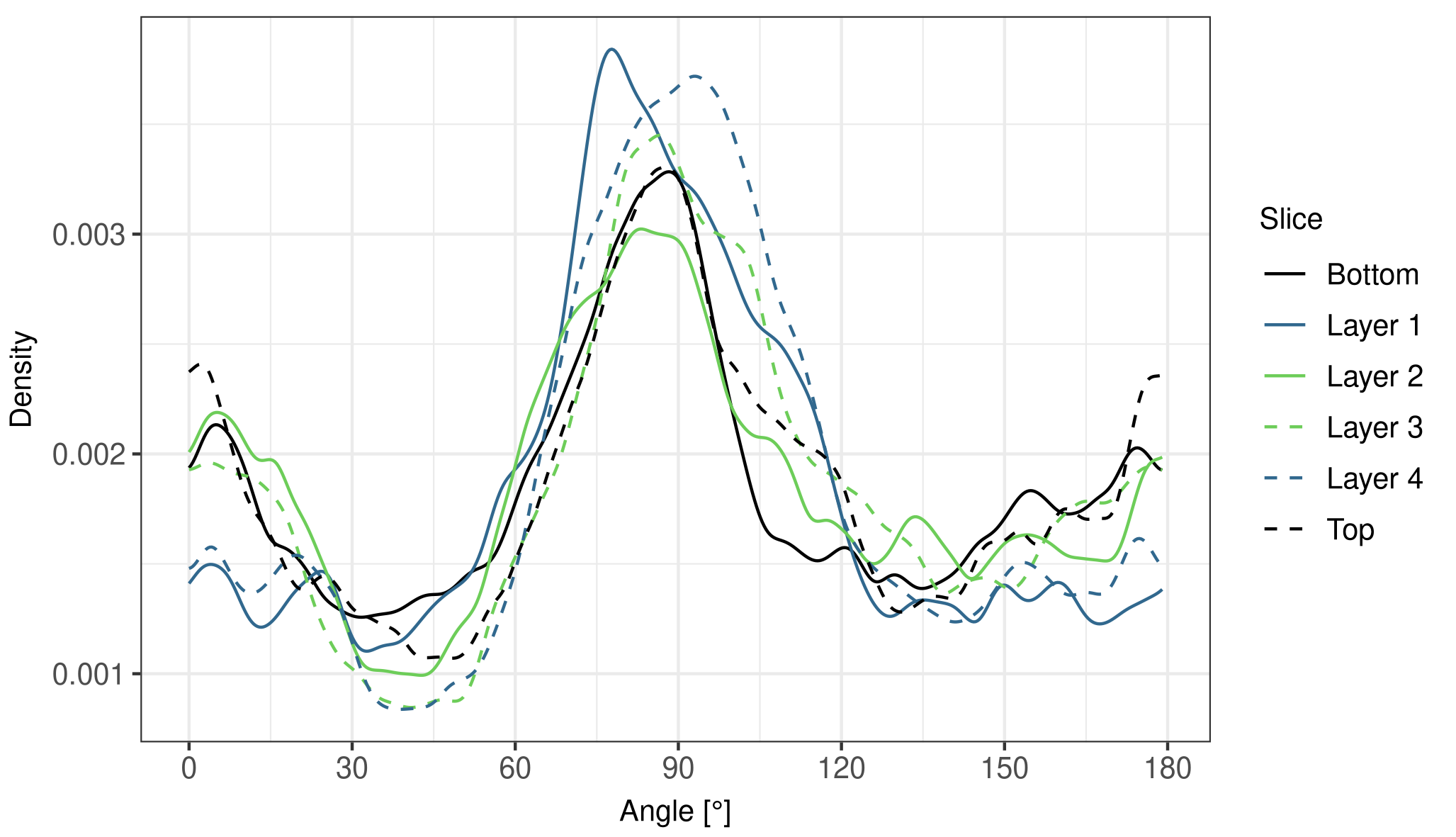}
\caption{Kernel density estimates for the fiber orientation distribution of specimen 19 (all 9 tiles combined), calculated from its \textmu CT scans.}\label{fig:densitiesSMC19_joint}
\end{figure}

\begin{figure}[!ht]%
\centering
\includegraphics[width=\textwidth]{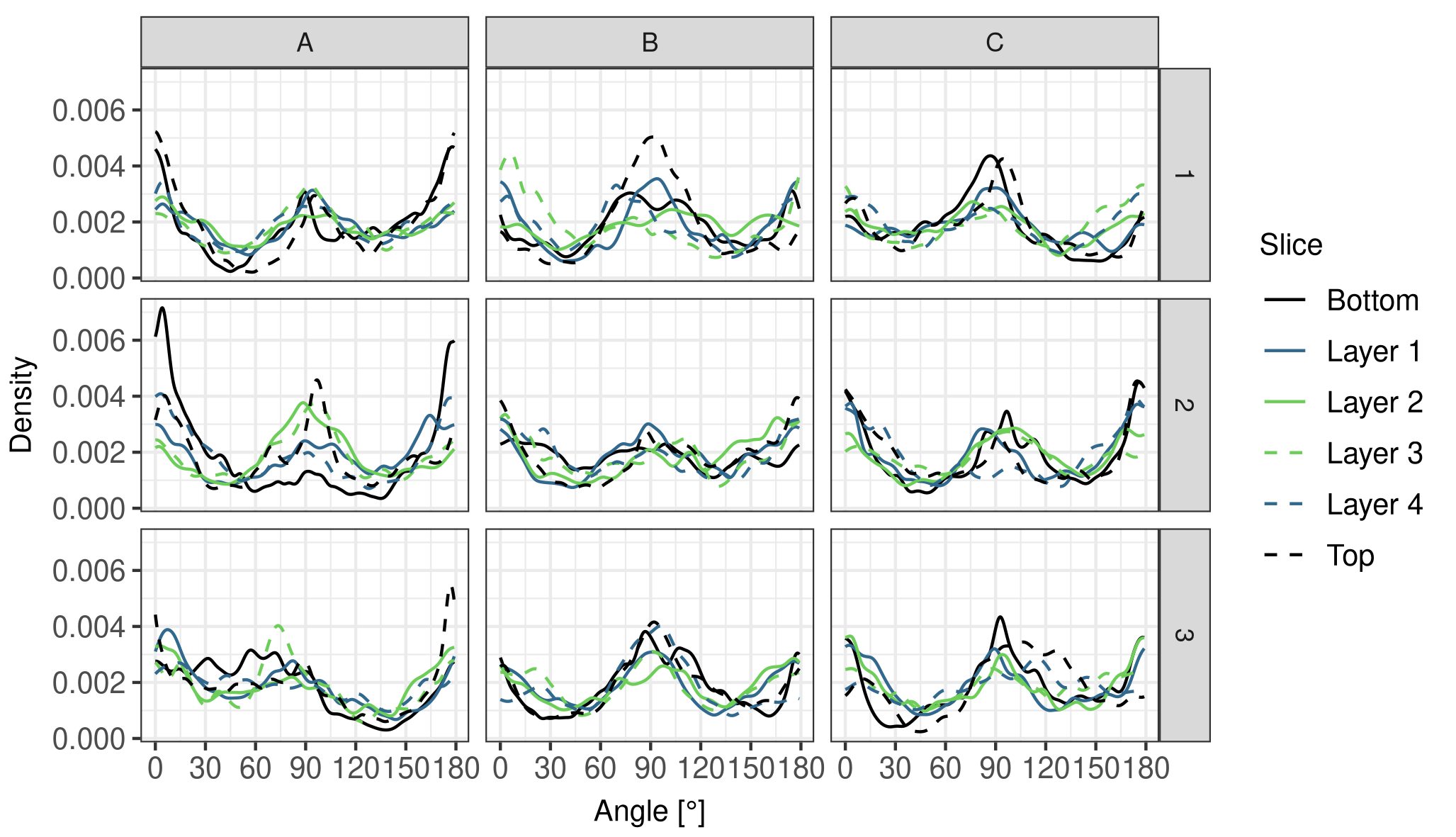}
\caption{Kernel density estimates for the fiber orientation distributions of the individual tiles of specimen 19, calculated from its \textmu CT scans.}\label{fig:densitiesSMC19_tiles}
\end{figure}

To address the question of how representative the surface layers are of the whole sample, we estimated the orientation densities for the surface layers and 4 interior layers. From Fig.~\ref{fig:densitiesSMC19_joint}, it can be seen that the fiber orientation distribution was considerably similar for both surface and interior layers. However, this might not be generalizable to all C-SMC materials as this behavior will depend on the production process and the size of the sample. For example, we chose the tile size slightly larger than the roving length. Thus, a tile's orientation histogram was dominated by the behavior of the few rovings that are observed in the field of view. So in this case, the surface layers were not representative of the whole sample, as can be seen in Fig.~\ref{fig:densitiesSMC19_tiles}. 
Further plots are provided in the supplementary material.

\section{Discussion}
\label{sec:discussion}

Imaging C-SMC is a challenging task due to the similar physics of carbon fibers and resin, resulting in low contrast images. Hence, polarization imaging is an attractive alternative as it can differentiate fibers and resin due to the fiber's polarization properties. Moreover, none of the established methods are applicable for inline quality inspection, whereas polarization imaging is fast, cheap, and yields a large field of view. Yet to our knowledge, it has only been validated for uni- and bidirectional carbon material. Since SMC material is multidirectional and, thus, more complex, we validated polarization imaging with CT imaging on C-SMC materials which are commercially available.

For the samples considered, we observed an encouraging visual similarity between both imaging modalities. However, we identified a considerable deviation between the polarization images and orientations estimated from CT images of up to $25^\circ$. Although computed tomography is an established method, its results should be taken with a grain of salt: Despite their high quality, the scans struggle with low contrast between fibers and resin. Remarkably, the MR method~\cite{robbFiberOrientationEstimation2007} succeeds in estimating fiber orientation, even though more commonly used methods for fiber orientation estimation fail~\cite{Keilmann_Godehardt_Moghiseh_Redenbach_Schladitz_2024}.

Nevertheless, even polarization imaging is not able to perfectly differentiate between fibers and resin: Atkinson et~al.~\cite{atkinson2018high} identified specular reflection of the resin as one possible cause of error. This might also explain the error range observed in this work: Resin-rich regions or regions where the carbon fibers are further below the surface of the specimen than others might be subject to larger errors due to the optical properties of the resin. In such areas, the polarized light ray and the angle of polarization might undergo some distortion as it passes through the resin before being registered by the camera's pixel sensor. This hypothesis could be validated by analyzing the local fiber area fraction in the slices and comparing it to the deviation plot in Fig.~\ref{fig:19t-error}. In addition, specimens of varying fiber volume fraction (in this work all specimens had an overall constant fiber weight fraction of 50\%) could be analyzed to prove or disprove this hypothesis.
Whereas the underlying mechanisms and sources of error for CT imaging are well studied~\cite{hsieh2003computed}, there exist only a few studies for polarization imaging~\cite{atkinson2018high, atkinson2021precision, Li:23, chiominto2024using}. Therefore, further work is planned to investigate its artifacts in more depth. 

Further research opportunities include the development of a fair comparison method between both imaging modalities: Since C-SMC surface layers are not perfectly planar, it was tricky to identify the uppermost layer in the CT scans. Still, the minimum over surface-adjacent layers yielded fairly good results. Moreover, image processing of angular data is rather rare, including metrics for image comparison. In this paper, we adapted the established mean absolute error (MAE) to the more appropriate mean angular deviance (MAD), taking into account the periodicity of angles. Although the MAE and the analogous mean squared error (MSE) are simple and well-explainable signal fidelity measures, they often fail to capture human perception of similarity~\cite{wang2009mean}. A similar, well-researched problem is the analysis of diffusion tensor images which also incorporate orientation information.
Such tensors also need to be normalized~\cite{munoz2009review, yangDiffusionTensorImage2008}. However, the difference measures used in this context are less intuitive than the MAE. Besides, different from angles, tensors provide information on the strength of orientation (anisotropy). Eventually, developing or adapting more sophisticated methods to angular data was outside the scope of this paper. 

Unlike computed tomography, polarization vision is limited to exterior layers. Therefore, we addressed the question of its representativity for the entire sample by developing a workflow to compare kernel density estimates of exterior and interior layers. For the examined specimens, the kernel estimates appear substantially similar. As this behavior depends on the composition and production process of the material, the representativity should be reevaluated for other materials.

\section{Conclusion}
We demonstrate the applicability of polarization imaging as a potential means of characterizing and analyzing the compression molding behavior of C-SMC materials complementary to CT scans. 
Polarization imaging of the surface and analysis of the orientation and degree of orientation images generated that way can be used to screen large numbers of specimens and even to monitor the C-SMC production process inline. 
For validation and to observe the inner layers inaccessible to polarization imaging, CT scans of a small subset of the specimens can complement the microstructural information.  
To demonstrate this approach, press rheometry characterization specimens of four types of commercially available C-SMC materials were imaged using polarization imaging and CT. 
Our analysis of fiber orientation using both methods demonstrates comparability of the surface fiber orientation data. It also indicates that for the current material selection, the surface fiber orientation distribution on the full specimen is representative of the fiber orientation distribution through the entire thickness of the material. Further work will include specimens with other material properties and 
seek to explain the considerable error we observed. This in turn will hopefully enable suitable corrections or better interpretation of the raw polarization image data.

\section*{Acknowledgments}
The authors thank Damjan Hatić for discussions about image registration. This work was supported by the German Federal Ministry of Education and Research under Grant Agreement No:~05M22UKA, and by the Fraunhofer~ITWM within the framework of the High Performance Center Simulation and Software Based Innovation as part of the project "C-SMC~Digitalization".

\section*{CRediT authorship contribution statement}
\textbf{Miro Duhovic:} Data curation, Formal analysis, Writing - original draft, Writing - review \& editing, Visualization. \textbf{Alex Keilmann:} Software, Data curation, Formal analysis, Methodology, Writing - original draft, Writing - review \& editing, Visualization. \textbf{Dominic Schommer:} Data curation, Formal analysis, Writing - review \& editing. \textbf{Claudia Redenbach:} Methodology, Supervision, Writing – review \& editing. \textbf{Katja Schladitz:} Conceptualization, Methodology, Project administration, Supervision, Writing – review \& editing

\section*{Code and Data Availability}
Our code for image registration and similarity measures (Section~\ref{sec:kernelDensity}) is available as jupyter notebook. The code for comparing kernel density estimates (Section~\ref{sec:kernelDensity}) of exterior and interior layers is available in R. Both are hosted together with our data on \url{10.5281/zenodo.13862633}. The code for the anisotropic Gaussian filters~\cite{Keilmann_Godehardt_Moghiseh_Redenbach_Schladitz_2024} used in the MR method is available at \url{github.com/akeilmann/anigauss}.

\bibliographystyle{IEEEtran}  
\bibliography{biblio}

\begin{thebibliography}{10}
\providecommand{\url}[1]{#1}
\csname url@samestyle\endcsname
\providecommand{\newblock}{\relax}
\providecommand{\bibinfo}[2]{#2}
\providecommand{\BIBentrySTDinterwordspacing}{\spaceskip=0pt\relax}
\providecommand{\BIBentryALTinterwordstretchfactor}{4}
\providecommand{\BIBentryALTinterwordspacing}{\spaceskip=\fontdimen2\font plus
\BIBentryALTinterwordstretchfactor\fontdimen3\font minus
  \fontdimen4\font\relax}
\providecommand{\BIBforeignlanguage}[2]{{%
\expandafter\ifx\csname l@#1\endcsname\relax
\typeout{** WARNING: IEEEtran.bst: No hyphenation pattern has been}%
\typeout{** loaded for the language `#1'. Using the pattern for}%
\typeout{** the default language instead.}%
\else
\language=\csname l@#1\endcsname
\fi
#2}}
\providecommand{\BIBdecl}{\relax}
\BIBdecl

\bibitem{fu96}
S.-Y. Fu and B.~Lauke, ``Effects of fiber length and fiber orientation
  distributions on the tensile strength of short-fiber-reinforced polymers,''
  \emph{Composites Science and Technology}, vol.~56, no.~10, pp. 1179--1190,
  1996.

\bibitem{vallons09}
K.~Vallons, I.~Duque, S.~Lomov, and I.~Verpoest, ``Fibre orientation effects on
  the tensile properties of biaxial carbon/epoxy ncf composites,'' in
  \emph{ICCM International Conferences on Composite Materials}, 01 2009, pp.
  27--31.

\bibitem{yokozeki06}
T.~Yokozeki, T.~Ogasawara, and T.~Ishikawa, ``Nonlinear behavior and
  compressive strength of unidirectional and multidirectional carbon fiber
  composite laminates,'' \emph{Composites Part A: Applied Science and
  Manufacturing}, vol.~37, no.~11, pp. 2069--2079, 2006.

\bibitem{heuer2015review}
H.~Heuer, M.~Schulze, M.~Pooch, S.~G{\"a}bler, A.~Nocke, G.~Bardl, C.~Cherif,
  M.~Klein, R.~Kupke, R.~Vetter \emph{et~al.}, ``Review on quality assurance
  along the cfrp value chain--non-destructive testing of fabrics, preforms and
  cfrp by hf radio wave techniques,'' \emph{Composites Part B: Engineering},
  vol.~77, pp. 494--501, 2015.

\bibitem{schommer2019material}
D.~Schommer, M.~Duhovic, V.~Romanenko, H.~Andr{\"a}, K.~Steiner, M.~Schneider,
  and J.~M. Hausmann, ``Material characterization and compression molding
  simulation of cf-smc materials in a press rheometry test,'' \emph{Key
  Engineering Materials}, vol. 809, pp. 467--472, 2019.

\bibitem{zhang11}
D.~Zhang, D.~E.~Smith, D.~A.~Jack, and S.~Montgomery-Smith, ``{Numerical
  Evaluation of Single Fiber Motion for Short-Fiber-Reinforced Composite
  Materials Processing},'' \emph{Journal of Manufacturing Science and
  Engineering}, vol. 133, no.~5, p. 051002, 09 2011.

\bibitem{denos2018fiber}
B.~R. Denos, D.~E. Sommer, A.~J. Favaloro, R.~B. Pipes, and W.~B. Avery,
  ``Fiber orientation measurement from mesoscale {CT} scans of prepreg platelet
  molded composites,'' \emph{Composites Part A: Applied Science and
  Manufacturing}, vol. 114, pp. 241--249, 2018.

\bibitem{garcea2018x}
S.~Garcea, Y.~Wang, and P.~Withers, ``X-ray computed tomography of polymer
  composites,'' \emph{Composites Science and Technology}, vol. 156, pp.
  305--319, 2018.

\bibitem{jensen2010directional}
T.~H. Jensen, M.~Bech, O.~Bunk, T.~Donath, C.~David, R.~Feidenhans, and
  F.~Pfeiffer, ``Directional x-ray dark-field imaging,'' \emph{Physics in
  Medicine \& Biology}, vol.~55, no.~12, p. 3317, 2010.

\bibitem{jensen2010directionalstrongly}
T.~H. Jensen, M.~Bech, I.~Zanette, T.~Weitkamp, C.~David, H.~Deyhle,
  S.~Rutishauser, E.~Reznikova, J.~Mohr, R.~Feidenhans’l \emph{et~al.},
  ``Directional x-ray dark-field imaging of strongly ordered systems,''
  \emph{Physical Review B—Condensed Matter and Materials Physics}, vol.~82,
  no.~21, p. 214103, 2010.

\bibitem{malecki2013x}
A.~D. Malecki, ``X-ray tensor tomography: From two-dimensional directional
  x-ray dark-field imaging to three dimensions,'' Ph.D. dissertation,
  Technische Universit{\"a}t M{\"u}nchen, 2013.

\bibitem{prade2017nondestructive}
F.~Prade, F.~Schaff, S.~Senck, P.~Meyer, J.~Mohr, J.~Kastner, and F.~Pfeiffer,
  ``Nondestructive characterization of fiber orientation in short fiber
  reinforced polymer composites with x-ray vector radiography,'' \emph{NDT \& E
  International}, vol.~86, pp. 65--72, 2017.

\bibitem{sharma2018advanced}
Y.~Sharma, ``Advanced acquisition methods for anisotropic x-ray dark-field
  imaging,'' Ph.D. dissertation, Technische Universit{\"a}t M{\"u}nchen, 2018.

\bibitem{schladitz2011quantitative}
K.~Schladitz, ``Quantitative micro-{CT},'' \emph{Journal of microscopy}, vol.
  243, no.~2, pp. 111--117, 2011.

\bibitem{bidola2017application}
P.~Bidola, K.~Morgan, M.~Willner, A.~Fehringer, S.~Allner, F.~Prade,
  F.~Pfeiffer, and K.~Achterhold, ``Application of sensitive, high-resolution
  imaging at a commercial lab-based x-ray micro-ct system using
  propagation-based phase retrieval,'' \emph{Journal of Microscopy}, vol. 266,
  no.~2, pp. 211--220, 2017.

\bibitem{jiang20203d}
N.~Jiang, Y.~Li, D.~Li, T.~Yu, Y.~Li, J.~Xu, N.~Li, and T.~J. Marrow, ``3d
  finite element modeling of water diffusion behavior of jute/pla composite
  based on x-ray computed tomography,'' \emph{Composites Science and
  Technology}, vol. 199, p. 108313, 2020.

\bibitem{latil2011towards}
P.~Latil, L.~Org{\'e}as, C.~Geindreau, P.~J. Dumont, and S.~R. Du~Roscoat,
  ``Towards the 3d in situ characterisation of deformation micro-mechanisms
  within a compressed bundle of fibres,'' \emph{Composites Science and
  Technology}, vol.~71, no.~4, pp. 480--488, 2011.

\bibitem{le2008XrayPhaseContrast}
T.-H. Le, P.~Dumont, L.~Orgéas, D.~Favier, L.~Salvo, and E.~Boller, ``X-ray
  phase contrast microtomography for the analysis of the fibrous microstructure
  of {{SMC}} composites,'' \emph{Composites Part A: Applied Science and
  Manufacturing}, vol.~39, no.~1, pp. 91--103, 2008.

\bibitem{martulli2019carbon}
L.~M. Martulli, L.~Muyshondt, M.~Kerschbaum, S.~Pimenta, S.~V. Lomov, and
  Y.~Swolfs, ``Carbon fibre sheet moulding compounds with high in-mould flow:
  Linking morphology to tensile and compressive properties,'' \emph{Composites
  Part A: Applied Science and Manufacturing}, vol. 126, p. 105600, 2019.

\bibitem{sabiston2019method}
T.~Sabiston, K.~Inal, and P.~Lee-Sullivan, ``Method to determine the required
  microstructure size to be represented by a second order fibre orientation
  tensor using x-ray micro computed tomography to evaluate compression moulded
  composites,'' \emph{Composites Science and Technology}, vol. 182, p. 107777,
  2019.

\bibitem{schladitzNondestructiveCharacterizationFiber2017}
K.~Schladitz, A.~Büter, M.~Godehardt, O.~Wirjadi, J.~Fleckenstein, T.~Gerster,
  U.~Hassler, K.~Jaschek, M.~Maisl, U.~Maisl, S.~Mohr, U.~Netzelmann,
  T.~Potyra, and M.~O. Steinhauser, ``Non-destructive characterization of fiber
  orientation in reinforced {{SMC}} as input for simulation based design,''
  \emph{Composite Structures}, vol. 160, pp. 195--203, 2017.

\bibitem{sentis20173d}
D.~F. Sentis, L.~Org{\'e}as, P.~J. Dumont, S.~R. Du~Roscoat, M.~Sager, and
  P.~Latil, ``3d in situ observations of the compressibility and pore transport
  in sheet moulding compounds during the early stages of compression
  moulding,'' \emph{Composites Part A: Applied Science and Manufacturing},
  vol.~92, pp. 51--61, 2017.

\bibitem{trauth2021effective}
A.~Trauth, L.~Kehrer, P.~Pinter, K.~Weidenmann, and T.~B{\"o}hlke, ``On the
  effective elastic properties based on mean-field homogenization of sheet
  molding compound composites,'' \emph{Composites Part C: Open Access}, vol.~4,
  p. 100089, 2021.

\bibitem{bardl2016automated}
G.~Bardl, A.~Nocke, C.~Cherif, M.~Pooch, M.~Schulze, H.~Heuer, M.~Schiller,
  R.~Kupke, and M.~Klein, ``Automated detection of yarn orientation in
  3d-draped carbon fiber fabrics and preforms from eddy current data,''
  \emph{Composites Part B: Engineering}, vol.~96, pp. 312--324, 2016.

\bibitem{de1992non}
M.~De~Goeje and K.~Wapenaar, ``Non-destructive inspection of carbon
  fibre-reinforced plastics using eddy current methods,'' \emph{Composites},
  vol.~23, no.~3, pp. 147--157, 1992.

\bibitem{lange1994structural}
R.~Lange and G.~Mook, ``Structural analysis of {CFRP} using eddy current
  methods,'' \emph{Ndt \& E International}, vol.~27, no.~5, pp. 241--248, 1994.

\bibitem{mook2001non}
G.~Mook, R.~Lange, and O.~Koeser, ``Non-destructive characterisation of
  carbon-fibre-reinforced plastics by means of eddy-currents,''
  \emph{Composites science and technology}, vol.~61, no.~6, pp. 865--873, 2001.

\bibitem{prakash1976eddy}
R.~Prakash and C.~Owston, ``Eddy-current method for the determination of lay-up
  order in cross-plied crfp laminates,'' \emph{Composites}, vol.~7, no.~2, pp.
  88--92, 1976.

\bibitem{yin2008noncontact}
W.~Yin, P.~J. Withers, U.~Sharma, and A.~J. Peyton, ``Noncontact
  characterization of carbon-fiber-reinforced plastics using multifrequency
  eddy current sensors,'' \emph{IEEE transactions on instrumentation and
  measurement}, vol.~58, no.~3, pp. 738--743, 2008.

\bibitem{zeng2019detection}
Z.~Zeng, J.~Wang, X.~Liu, J.~Lin, and Y.~Dai, ``Detection of fiber waviness in
  cfrp using eddy current method,'' \emph{Composite Structures}, vol. 229, p.
  111411, 2019.

\bibitem{romanenko2020materialcharakterisierung}
V.~Romanenko, \emph{Materialcharakterisierung und durchg{\"a}ngige
  3D-Prozesssimulation f{\"u}r kohlenstofffaserverst{\"a}rktes Sheet Molding
  Compound}.\hskip 1em plus 0.5em minus 0.4em\relax Technische Universit{\"a}t
  Kaiserslautern, 2020.

\bibitem{Sukhanov_2019}
\BIBentryALTinterwordspacing
D.~Sukhanov, K.~Zavyalova, and A.~Kadurina, ``Method for enhancement of spatial
  resolution of eddy current imaging,'' \emph{Measurement Science and
  Technology}, vol.~30, no.~6, p. 065402, apr 2019. [Online]. Available:
  \url{dx.doi.org/10.1088/1361-6501/ab0b10}
\BIBentrySTDinterwordspacing

\bibitem{suragus22}
\BIBentryALTinterwordspacing
{Suragus~Website}, {A}ccessed on 31.10.2022. [Online]. Available:
  \url{suragus.com/media/filer_public/18/6e/186e9c22-d4eb-4f21-9cc3-04db9927f1b1/flyer_cf_sensor_kit_for_robot_integration_v10_web.pdf}
\BIBentrySTDinterwordspacing

\bibitem{suragus23}
\BIBentryALTinterwordspacing
------, {A}ccessed on 02.03.2023. [Online]. Available:
  \url{suragus.com/de/produkte/karbonfaser-pruefung/inline-systeme/roboter-fuer-3d-carbonfaser-pruefung/}
\BIBentrySTDinterwordspacing

\bibitem{aindow1986fibre}
J.~Aindow, M.~Markham, K.~Puttick, J.~Rider, and M.~Rudman, ``Fibre orientation
  detection in injection-moulded carbon fibre reinforced components by
  thermography and ultrasonics,'' \emph{NDT international}, vol.~19, no.~1, pp.
  24--29, 1986.

\bibitem{karpen1994depth}
W.~Karpen, D.~Wu, R.~Steegmuller, and G.~Busse, ``Depth profiling of
  orientation in laminates with local lockin thermography,'' in \emph{Proc.
  QIRT}, vol.~94, 1994, pp. 23--26.

\bibitem{fernandes2013use}
H.~Fernandes and X.~Maldague, ``Use of infrared thermography to measure fiber
  orientation on carbon-fiber reinforced composites,'' in \emph{Proceedings of
  the 16th International Symposium on Applied Electromagnetics and Mechanics
  (ISEM), in Quebec City}, 2013.

\bibitem{eberhardt2001fibre}
C.~Eberhardt and A.~Clarke, ``Fibre-orientation measurements in
  short-glass-fibre composites. {P}art {I}: automated, high-angular-resolution
  measurement by confocal microscopy,'' \emph{Composites Science and
  Technology}, vol.~61, no.~10, pp. 1389--1400, 2001.

\bibitem{hayes10Optical}
B.~S. Hayes, ``Optical microscopy of composites,''
  \url{app.knovel.com/hotlink/toc/id:kpOMFRC00F/optical-microscopy-fiber/optical-microscopy-fiber},
  2010, [Online].

\bibitem{lee2003measurement}
K.~S. Lee, S.~W. Lee, K.~Chung, T.~J. Kang, and J.~R. Youn, ``Measurement and
  numerical simulation of three-dimensional fiber orientation states in
  injection-molded short-fiber-reinforced plastics,'' \emph{Journal of applied
  polymer science}, vol.~88, no.~2, pp. 500--509, 2003.

\bibitem{textechno22}
{Textechno~Website},
  \url{textechno.com/de/internationale-standardisierung-der-testmethode-des-drapetest/},
  2022, {A}ccessed on 31.10.2022.

\bibitem{profactor2}
{ProFactor~Website},
  \url{profactor.at/wp-content/uploads/2019/02/PROFACTOR_ZeroDefectManufacturing_final_web.pdf},
  2022, {A}ccessed on 31.10.2022.

\bibitem{profactor1}
------,
  \url{profactor.at/loesungen/bildverarbeitung-machine-vision/faserverbundwerkstoffe/},
  2022, {A}ccessed on 31.10.2022.

\bibitem{cikoni22}
{CIKONI~Website},
  \url{cikoni.com/de/das-robotergestuetzte-qs-system-drapewatch-ueberwacht-ab-
  jetzt-die-fertigung\ -von-carbonbauteile-in-der-fabrik-der-zukunft}, 2022,
  {A}ccessed on 31.10.2022.

\bibitem{swinburne22}
{Swinburne~University~of~Technology~Website},
  \url{swinburne.edu.au/research/platforms-initiatives/factory-of-the-future/},
  2022, {A}ccessed on 31.10.2022.

\bibitem{hexagon22}
{Hexagon~Web~Site}, ``Pioneering composite inspection technology,''
  \url{hexagon.com/resources/resource-library/pioneering-composite-inspection-technology},
  2022, {A}ccessed on 31.10.2022.

\bibitem{ccMagazin18}
D.~Franke, ``Durchgängige {Q}ualität sichert {E}rfolg,'' \emph{Carbon
  Composites Magazin}, vol.~1, 2018.

\bibitem{dobrich2023machine}
O.~D{\"o}brich and C.~Brauner, ``Machine vision system for digital twin
  modeling of composite structures,'' \emph{Frontiers in Materials}, vol.~10,
  p. 1154655, 2023.

\bibitem{freitag2015theoretical}
\BIBentryALTinterwordspacing
C.~Freitag, L.~Alter, R.~Weber, T.~Grafa, and T.~Graf, ``Theoretical and
  experimental determination of the polarization dependent absorptance of laser
  radiation in carbon fibers and cfrp,'' \emph{Lasers in Manufacturing.}, 2015.
  [Online]. Available:
  \url{www.wlt.de/lim/Proceedings2015/Stick/PDF/Contribution111_final.pdf}
\BIBentrySTDinterwordspacing

\bibitem{stokes1851composition}
G.~G. Stokes, ``On the composition and resolution of streams of polarized light
  from different sources,'' \emph{Transactions of the Cambridge Philosophical
  Society}, vol.~9, p. 399, 1851.

\bibitem{urabe1992rotative}
K.~Urabe, ``Rotative polarization system of millimetric wave for detecting
  fiber orientation in cfrp,'' \emph{Journal of reinforced plastics and
  composites}, vol.~11, no.~2, pp. 179--197, 1992.

\bibitem{urabe1991nondestructive}
K.~Urabe and S.~Yomoda, ``A nondestructive testing method of fiber orientation
  by microwave,'' \emph{Advanced Composite Materials}, vol.~1, no.~3, pp.
  193--208, 1991.

\bibitem{gulhane2023advance}
P.~D. Gulhane, S.~G. Jawarkar, Y.~Belsare, and S.~Bhalerao, ``Advance and
  review of composite material non-destructive testing techniques,'' \emph{Int.
  J. Creat. Res. Thoughts}, vol.~11, pp. 2320--2882, 2023.

\bibitem{schoberl2016measuring}
\BIBentryALTinterwordspacing
M.~Sch{\"o}berl, K.~Kasnakli, and A.~Nowak, ``Measuring strand orientation in
  carbon fiber reinforced plastics (cfrp) with polarization,'' \emph{World
  conference on non-destructive testing.}, 2016. [Online]. Available:
  \url{www.ndt.net/article/wcndt2016/papers/p20.pdf}
\BIBentrySTDinterwordspacing

\bibitem{wang2020non}
B.~Wang, S.~Zhong, T.-L. Lee, K.~S. Fancey, and J.~Mi, ``Non-destructive
  testing and evaluation of composite materials/structures: A state-of-the-art
  review,'' \emph{Advances in mechanical engineering}, vol.~12, no.~4, p.
  1687814020913761, 2020.

\bibitem{atkinson2018high}
G.~A. Atkinson and J.~D. Ernst, ``High-sensitivity analysis of polarization by
  surface reflection,'' \emph{Machine Vision and Applications}, vol.~29, pp.
  1171--1189, 2018.

\bibitem{ernst2014messung}
J.~Ernst, S.~Junger, and W.~Tschekalinskij, ``Messung einer {F}aserrichtung
  eines {K}ohlefaserwerkstoffes und {H}erstellung eines {O}bjekts in
  {K}ohlefaserverbundbauweise,'' 2014, dEPatent, 10 2012 220 923.9,.

\bibitem{ernst2016measurement}
------, ``Measurement of a fiber direction of a carbon fiber material and
  fabrication of an object in carbon fiber composite technique,'' Jan.~12 2016,
  uS Patent 9,234,836.

\bibitem{duhovic2022digitizing}
M.~Duhovic, T.~Hoffmann, D.~Schommer, J.~Ernst, K.~Schladitz, A.~Moghiseh,
  F.~Gortner, J.~M. Hausmann, P.~Mitschang, and K.~Steiner, ``Digitizing the
  production of carbon fiber sheet molding compounds,'' 2022.

\bibitem{schommer2023polarization}
D.~Schommer, M.~Duhovic, T.~Hoffmann, J.~Ernst, K.~Schladitz, A.~Moghiseh,
  F.~Gortner, J.~Hausmann, P.~Mitschang, and K.~Steiner, ``Polarization imaging
  for surface fiber orientation measurements of carbon fiber sheet molding
  compounds,'' \emph{Composites Communications}, vol.~37, p. 101456, 2023.

\bibitem{shen2023}
G.~Shen, R.~Ye, M.~Li, Z.~Liu, Z.~Ying, S.~Li, X.~Wang, L.~Li, M.~Zhou,
  H.~Zhou, and Y.~Zhang, ``Decoupling polymer molecular chains and carbon
  fibres orientation by their dielectric anisotropy,'' \emph{Composites Part A:
  Applied Science and Manufacturing}, vol. 175, p. 107768, 2023.

\bibitem{atkinson2021precision}
G.~A. Atkinson, S.~O’Hara~Nash, and L.~N. Smith, ``Precision fibre angle
  inspection for carbon fibre composite structures using polarisation vision,''
  \emph{Electronics}, vol.~10, no.~22, p. 2765, 2021.

\bibitem{chiominto2024using}
L.~Chiominto, G.~D’Emilia, and E.~Natale, ``Using light polarization to
  identify fiber orientation in carbon fiber components: metrological
  analysis,'' 2024.

\bibitem{SMCarbon3k}
{Polynt~Composites~Germany~GmbH}, ``{Polynt Composites: Sicherheitsdatenblatt
  SMCarbon}\textsuperscript \textregistered 24 {CF50-3k},'' 2016.

\bibitem{SMCarbon3kt}
------, ``Polynt {Composites}: {Technical Datasheet SMCarbon}\textsuperscript
  \textregistered 24 {CF50-3K},'' www.polynt.com, 2016.

\bibitem{CompoundsCarbonFiber}
------, ``{Polynt Composites: Compounds Carbon Fiber},'' 2019.

\bibitem{SMCarbon12k}
------, ``{Polynt Composites: Sicherheitsdatenblatt SMCarbon}\textsuperscript
  \textregistered 24 {CF50-12k},'' 2016.

\bibitem{SMCarbon12kt}
------, ``{Polynt Composites: Technical Datasheet SMCarbon}\textsuperscript
  \textregistered 24 {CF50-12K},'' 2016.

\bibitem{AMC85593}
N.A., ``{A. Schulman: Safety Data Sheet AMC} 85593,'' 2018.

\bibitem{AMC85590}
------, ``{A. Schulman: Safety Data Sheet AMC} 85590,'' 2018.

\bibitem{AMC8593t}
------, ``{Quantum Composites: Technical Data Sheet AMC}\textsuperscript
  \textregistered 8593,'' 2018.

\bibitem{AMC8590t}
------, ``{Quantum Composites: Technical Data Sheet AMC}\textsuperscript
  \textregistered 8590,'' 2018.

\bibitem{romanenko2022process-simulation}
V.~Romanenko, M.~Duhovic, D.~Schommer, J.~Hausmann, and J.~Eschl, ``Advanced
  process simulation of compression molded carbon fiber sheet molding compound
  (c-smc) parts in automotive series applications,'' \emph{Composites Part A:
  Applied Science and Manufacturing}, vol. 157, p. 106924, 2022.

\bibitem{robbFiberOrientationEstimation2007}
K.~Robb, O.~Wirjadi, and K.~Schladitz, ``Fiber {{Orientation Estimation}} from
  {{3D Image Data}}: {{Practical Algorithms}}, {{Visualization}}, and
  {{Interpretation}},'' in \emph{7th {{International Conference}} on {{Hybrid
  Intelligent Systems}} ({{HIS}} 2007)}.\hskip 1em plus 0.5em minus 0.4em\relax
  {IEEE}, 2007, pp. 320--325.

\bibitem{mavi}
{Fraunhofer~ITWM}, ``{MAVI} - modular algorithms for volume images,''
  \url{www.mavi-3d.de}, 2022, accessed: 2022-09-20.

\bibitem{toolip}
------, ``{ToolIP} tool for image processing,''
  \url{www.itwm.fraunhofer.de/toolip}, 2022, accessed: 2022-09-20.

\bibitem{Keilmann_Godehardt_Moghiseh_Redenbach_Schladitz_2024}
A.~Keilmann, M.~Godehardt, A.~Moghiseh, C.~Redenbach, and K.~Schladitz,
  ``Improved anisotropic gaussian filters,'' \emph{Image Analysis and
  Stereology}, vol.~43, no.~1, p. 9–22, 2024.

\bibitem{niblack1985introduction}
W.~Niblack, \emph{An introduction to digital image processing}.\hskip 1em plus
  0.5em minus 0.4em\relax Strandberg Publishing Company, 1985.

\bibitem{gonzalez2018DigitalImageProcessing}
R.~C. Gonzalez and R.~E. Woods, \emph{Digital Image Processing}.\hskip 1em plus
  0.5em minus 0.4em\relax Pearson, 2018.

\bibitem{mardia2000DirectionalStatistics}
K.~V. Mardia and P.~E. Jupp, \emph{Directional Statistics}, ser. Wiley Series
  in Probability and Statistics.\hskip 1em plus 0.5em minus 0.4em\relax {J.
  Wiley}, 2000.

\bibitem{sheather91}
S.~J. Sheather and M.~C. Jones, ``A reliable data-based bandwidth selection
  method for kernel density estimation,'' \emph{Journal of the Royal
  Statistical Society: Series B (Methodological)}, vol.~53, no.~3, pp.
  683--690, 1991.

\bibitem{hsieh2003computed}
J.~Hsieh, \emph{Computed tomography: principles, design, artifacts, and recent
  advances}.\hskip 1em plus 0.5em minus 0.4em\relax SPIE press, 2003.

\bibitem{Li:23}
L.~W. Li, N.~A. Rubin, M.~Juhl, J.-S. Park, and F.~Capasso, ``Evaluation and
  characterization of imaging polarimetry through metasurface polarization
  gratings,'' \emph{Appl. Opt.}, vol.~62, no.~7, pp. 1704--1722, 2023.

\bibitem{wang2009mean}
Z.~Wang and A.~C. Bovik, ``Mean squared error: Love it or leave it? a new look
  at signal fidelity measures,'' \emph{IEEE signal processing magazine},
  vol.~26, no.~1, pp. 98--117, 2009.

\bibitem{munoz2009review}
E.~Mu{\~n}oz-Moreno, R.~C{\'a}rdenes-Almeida, and M.~Martin-Fernandez, ``Review
  of techniques for registration of diffusion tensor imaging,'' \emph{Tensors
  in Image Processing and Computer Vision}, pp. 273--297, 2009.

\bibitem{yangDiffusionTensorImage2008}
J.~Yang, D.~Shen, C.~Davatzikos, and R.~Verma, ``Diffusion {{Tensor Image
  Registration Using Tensor Geometry}} and {{Orientation Features}},'' in
  \emph{Medical {{Image Computing}} and {{Computer-Assisted Intervention}} –
  {{MICCAI}} 2008}, D.~Metaxas, L.~Axel, G.~Fichtinger, and G.~Székely,
  Eds.\hskip 1em plus 0.5em minus 0.4em\relax {Springer Berlin Heidelberg},
  2008, vol. 5242, pp. 905--913.

\end{thebibliography}
\end{document}